\documentclass[aps,prl,reprint,longbibliography,preprintnumbers,floatfix,superscriptaddress]{revtex4-2}
\usepackage{graphicx} 
\usepackage{dcolumn}
\usepackage{bm}
\usepackage{amsmath}
\usepackage{multirow}
\usepackage{placeins} 
\usepackage{soul}
\usepackage{booktabs} 
\usepackage{xspace, tabularx}
\usepackage[caption=false]{subfig}
\usepackage{accents}

\usepackage{threeparttable}
\usepackage{lineno}
\usepackage{siunitx}

\makeatletter
\newcommand{\oset}[3][0ex]{%
  \mathrel{\mathop{#3}\limits^{
    \vbox to#1{\kern-2.5\ex@
    \hbox{$\scriptstyle#2$}\vss}}}}
\makeatother

\newcommand{\nue}{\ensuremath{\nu_{e}}\xspace}

\newcommand{\nuebar}{\ensuremath{\bar{\nu}_e}\xspace}

\newcommand{\nueParen}{\ensuremath{{\nu}_{e}~(\bar{\nu}_{e})}\xspace}

\newcommand{\numu}{\ensuremath{\nu_{\mu}}\xspace}

\newcommand{\numubar}{\ensuremath{\bar{\nu}_\mu}\xspace}

\newcommand{\numuParen}{\ensuremath{{\nu}_{\mu}~(\bar{\nu}_{\mu})}\xspace}

\newcommand{\antinumu}{\numubar}
\newcommand{\nutau}{\ensuremath{\nu_{\tau}}\xspace}

\newcommand{\nuone}{\ensuremath{\nu_{1}}\xspace}

\newcommand{\nutwo}{\ensuremath{\nu_{2}}\xspace}

\newcommand{\nuthree}{\ensuremath{\nu_{3}}\xspace}

\newcommand{\dmsq}{\ensuremath{\Delta m^2_{32}}\xspace}
\newcommand{\snsq}{\ensuremath{\sin^2 \theta_{23}}\xspace}
\newcommand{\thatm}{\ensuremath{\theta_{23}}\xspace}
\newcommand{\deltacp}{\ensuremath{\delta_{\rm{CP}}}\xspace}

\newcommand{\dmsqtwoone}{\ensuremath{\Delta m^2_{21}}\xspace}

\newcommand{\ppfx}{\textsc{PPFX}}
\newcommand{\geant}{\textsc{Geant4}\xspace}
\newcommand{\menate}{\textsc{Menate}\xspace}
\newcommand{\genie}{\textsc{Genie}\xspace}
\newcommand{\valencia}{Val\`{e}ncia}
\newcommand{\geantrwgt}{\textsc{Geant4Reweight}\xspace}

\newcommand{\timess}{$\times$}

\newcommand{\papertitle}{Precision measurement of neutrino oscillation parameters with 10 years of data from the NOvA experiment}

\usepackage{subfiles} 

\begin{document}

\title{\papertitle}
\preprint{FERMILAB-PUB-25-0619-PPD} 

\newcommand{\ANL}{Argonne National Laboratory, Argonne, Illinois 60439, 
USA}
\newcommand{\Bandirma}{Bandirma Onyedi Eyl\"ul University, Faculty of 
Engineering and Natural Sciences, Engineering Sciences Department, 
10200, Bandirma, Balıkesir, Turkey}
\newcommand{\ICS}{Institute of Computer Science, The Czech 
Academy of Sciences, 
182 07 Prague, Czech Republic}
\newcommand{\IOP}{Institute of Physics, The Czech 
Academy of Sciences, 
182 21 Prague, Czech Republic}
\newcommand{\Atlantico}{Universidad del Atlantico,
Carrera 30 No.\ 8-49, Puerto Colombia, Atlantico, Colombia}
\newcommand{\BHU}{Department of Physics, Institute of Science, Banaras 
Hindu University, Varanasi, 221 005, India}
\newcommand{\UCLA}{Physics and Astronomy Department, UCLA, Box 951547, Los 
Angeles, California 90095-1547, USA}
\newcommand{\Caltech}{California Institute of 
Technology, Pasadena, California 91125, USA}
\newcommand{\Cochin}{Department of Physics, Cochin University
of Science and Technology, Kochi 682 022, India}
\newcommand{\Charles}
{Charles University, Faculty of Mathematics and Physics,
 Institute of Particle and Nuclear Physics, Prague, Czech Republic}
\newcommand{\Cincinnati}{Department of Physics, University of Cincinnati, 
Cincinnati, Ohio 45221, USA}
\newcommand{\CSU}{Department of Physics, Colorado 
State University, Fort Collins, CO 80523-1875, USA}
\newcommand{\CTU}{Czech Technical University in Prague,
Brehova 7, 115 19 Prague 1, Czech Republic}
\newcommand{\Dallas}{Physics Department, University of Texas at Dallas,
800 W. Campbell Rd. Richardson, Texas 75083-0688, USA}
\newcommand{\DallasU}{University of Dallas, 1845 E 
Northgate Drive, Irving, Texas 75062 USA}
\newcommand{\Delhi}{Department of Physics and Astrophysics, University of 
Delhi, Delhi 110007, India}
\newcommand{\JINR}{Joint Institute for Nuclear Research,  
Dubna, Moscow region 141980, Russia}
\newcommand{\Erciyes}{
Department of Physics, Erciyes University, Kayseri 38030, Turkey}
\newcommand{\FNAL}{Fermi National Accelerator Laboratory, Batavia, 
Illinois 60510, USA}
\newcommand{\FSU}{Florida State University, Tallahassee, Florida 32306, USA}
\newcommand{\UFG}{Instituto de F\'{i}sica, Universidade Federal de 
Goi\'{a}s, Goi\^{a}nia, Goi\'{a}s, 74690-900, Brazil}
\newcommand{\Guwahati}{Department of Physics, IIT Guwahati, Guwahati, 781 
039, India}
\newcommand{\Harvard}{Department of Physics, Harvard University, 
Cambridge, Massachusetts 02138, USA}
\newcommand{\Houston}{Department of Physics, 
University of Houston, Houston, Texas 77204, USA}
\newcommand{\IHyderabad}{Department of Physics, IIT Hyderabad, Hyderabad, 
502 205, India}
\newcommand{\Hyderabad}{School of Physics, University of Hyderabad, 
Hyderabad, 500 046, India}
\newcommand{\IIT}{Illinois Institute of Technology,
Chicago IL 60616, USA}
\newcommand{\Indiana}{Indiana University, Bloomington, Indiana 47405, 
USA}
\newcommand{\INR}{Institute for Nuclear Research of Russia, Academy of 
Sciences 7a, 60th October Anniversary prospect, Moscow 117312, Russia}
\newcommand{\UIowa}{Department of Physics and Astronomy, University of Iowa, 
Iowa City, Iowa 52242, USA}
\newcommand{\Iowa}{Department of Physics and Astronomy, Iowa State 
University, Ames, Iowa 50011, USA}
\newcommand{\Irvine}{Department of Physics and Astronomy, 
University of California at Irvine, Irvine, California 92697, USA}
\newcommand{\Jammu}{Department of Physics and Electronics, University of 
Jammu, Jammu Tawi, 180 006, Jammu and Kashmir, India}
\newcommand{\Lebedev}{Nuclear Physics and Astrophysics Division, Lebedev 
Physical 
Institute, Leninsky Prospect 53, 119991 Moscow, Russia}
\newcommand{\Magdalena}{Universidad del Magdalena, Carrera 32 No 22-08 Santa Marta, Colombia}
\newcommand{\MSU}{Department of Physics and Astronomy, Michigan State 
University, East Lansing, Michigan 48824, USA}
\newcommand{\Crookston}{Math, Science and Technology Department, University 
of Minnesota Crookston, Crookston, Minnesota 56716, USA}
\newcommand{\Duluth}{Department of Physics and Astronomy, 
University of Minnesota Duluth, Duluth, Minnesota 55812, USA}
\newcommand{\Minnesota}{School of Physics and Astronomy, University of 
Minnesota Twin Cities, Minneapolis, Minnesota 55455, USA}
\newcommand{\Mississippi}{University of Mississippi, University, Mississippi 38677, USA}
\newcommand{\NISER}{National Institute of Science Education and Research, An OCC of Homi Bhabha
National Institute, Bhubaneswar, Odisha, India}
\newcommand{\OSU}{Department of Physics, Ohio State University, Columbus,
Ohio 43210, USA}
\newcommand{\Oxford}{Subdepartment of Particle Physics, 
University of Oxford, Oxford OX1 3RH, United Kingdom}
\newcommand{\Panjab}{Department of Physics, Panjab University, 
Chandigarh, 160 014, India}
\newcommand{\Pitt}{Department of Physics, 
University of Pittsburgh, Pittsburgh, Pennsylvania 15260, USA}
\newcommand{\QMU}{Particle Physics Research Centre, 
Department of Physics and Astronomy,
Queen Mary University of London,
London E1 4NS, United Kingdom}
\newcommand{\RAL}{Rutherford Appleton Laboratory, Science 
and 
Technology Facilities Council, Didcot, OX11 0QX, United Kingdom}
\newcommand{\SAlabama}{Department of Physics, University of 
South Alabama, Mobile, Alabama 36688, USA} 
\newcommand{\Carolina}{Department of Physics and Astronomy, University of 
South Carolina, Columbia, South Carolina 29208, USA}
\newcommand{\SDakota}{South Dakota School of Mines and Technology, Rapid 
City, South Dakota 57701, USA}
\newcommand{\SMU}{Department of Physics, Southern Methodist University, 
Dallas, Texas 75275, USA}
\newcommand{\Stanford}{Department of Physics, Stanford University, 
Stanford, California 94305, USA}
\newcommand{\Sussex}{Department of Physics and Astronomy, University of 
Sussex, Falmer, Brighton BN1 9QH, United Kingdom}
\newcommand{\Syracuse}{Department of Physics, Syracuse University,
Syracuse NY 13210, USA}
\newcommand{\Tennessee}{Department of Physics and Astronomy, 
University of Tennessee, Knoxville, Tennessee 37996, USA}
\newcommand{\Texas}{Department of Physics, University of Texas at Austin, 
Austin, Texas 78712, USA}
\newcommand{\Tufts}{Department of Physics and Astronomy, Tufts University, Medford, 
Massachusetts 02155, USA}
\newcommand{\UCL}{Physics and Astronomy Department, University College 
London, 
Gower Street, London WC1E 6BT, United Kingdom}
\newcommand{\Virginia}{Department of Physics, University of Virginia, 
Charlottesville, Virginia 22904, USA}
\newcommand{\WSU}{Department of Mathematics, Statistics, and Physics,
 Wichita State University, 
Wichita, Kansas 67260, USA}
\newcommand{\WandM}{Department of Physics, William \& Mary, 
Williamsburg, Virginia 23187, USA}
\newcommand{\Wisconsin}{Department of Physics, University of 
Wisconsin-Madison, Madison, Wisconsin 53706, USA}
\newcommand{\deceased}{Deceased.}
\affiliation{\ANL}
\affiliation{\Atlantico}
\affiliation{\Bandirma}
\affiliation{\BHU}
\affiliation{\Caltech}
\affiliation{\Charles}
\affiliation{\Cincinnati}
\affiliation{\Cochin}
\affiliation{\CSU}
\affiliation{\CTU}
\affiliation{\Delhi}
\affiliation{\Erciyes}
\affiliation{\FNAL}
\affiliation{\FSU}
\affiliation{\UFG}
\affiliation{\Guwahati}
\affiliation{\Houston}
\affiliation{\Hyderabad}
\affiliation{\IHyderabad}
\affiliation{\IIT}
\affiliation{\Indiana}
\affiliation{\ICS}
\affiliation{\INR}
\affiliation{\IOP}
\affiliation{\Iowa}
\affiliation{\Irvine}
\affiliation{\JINR}
\affiliation{\Magdalena}
\affiliation{\MSU}
\affiliation{\Duluth}
\affiliation{\Minnesota}
\affiliation{\Mississippi}
\affiliation{\OSU}
\affiliation{\NISER}
\affiliation{\Panjab}
\affiliation{\Pitt}
\affiliation{\QMU}
\affiliation{\SAlabama}
\affiliation{\Carolina}
\affiliation{\SMU}
\affiliation{\Sussex}
\affiliation{\Syracuse}
\affiliation{\Texas}
\affiliation{\Tufts}
\affiliation{\UCL}
\affiliation{\Virginia}
\affiliation{\WSU}
\affiliation{\WandM}
\affiliation{\Wisconsin}

\author{S.~Abubakar}
\affiliation{\Erciyes}

\author{M.~A.~Acero}
\affiliation{\Atlantico}

\author{B.~Acharya}
\affiliation{\Mississippi}

\author{P.~Adamson}
\affiliation{\FNAL}









\author{N.~Anfimov}
\affiliation{\JINR}


\author{A.~Antoshkin}
\affiliation{\JINR}


\author{E.~Arrieta-Diaz}
\affiliation{\Magdalena}

\author{L.~Asquith}
\affiliation{\Sussex}


\author{A.~Aurisano}
\affiliation{\Cincinnati}


\author{D.~Azevedo}
\affiliation{\UFG}

% ...

\collaboration{The NOvA Collaboration}
\noaffiliation




\begin{abstract}
This Letter reports measurements of muon-neutrino disappearance and electron-neutrino appearance and the corresponding antineutrino processes between the two NOvA detectors in the NuMI neutrino beam. These measurements use a dataset with double the neutrino mode beam exposure that was previously analyzed, along with improved simulation and analysis techniques. A joint fit to these samples in the three-flavor paradigm results in the most precise single-experiment constraint on the atmospheric neutrino mass splitting, $\Delta m^2_{32}= 2.431^{+0.036}_{-0.034} (-2.479^{+0.036}_{-0.036}) \times 10^{-3}~\mathrm{eV}^2$ if the mass ordering is normal (inverted). In both orderings, a region close to maximal mixing with $\sin^2 \theta_{23}=0.55^{+0.02}_{-0.06}$ is preferred. The NOvA data show a mild preference for the normal mass ordering with a Bayes factor of 2.4 (corresponding to 70\% of the posterior probability), indicating that the normal ordering is 2.4 times more probable than the inverted ordering. When incorporating a 2D $\Delta m^2_{32}$\textendash$\sin^2 2\theta_{13}$  constraint based on Daya Bay data, this preference strengthens to a Bayes factor of 6.6 (87\%).
\end{abstract}

\maketitle

Neutrino oscillations~\cite{pontecorvo1968, Pontekorvo1978} describe the phenomenon by which a neutrino produced in a particular flavor at one point in spacetime can subsequently be detected as a different flavor. This flavor transformation occurs due to mixing~\cite{Maki:1962mu} between the flavor eigenstates (\nue, \numu, \nutau) and the mass eigenstates (\nuone, \nutwo, \nuthree). For a unitary mixing matrix, neutrino oscillation probabilities can be characterized by three mixing angles ($\theta_{12}$, $\theta_{13}$, \thatm), the charge-parity (CP) violation phase ($\delta_{\rm CP}$), and two nonzero mass splittings (\dmsqtwoone and \dmsq, where $\Delta m^2_{ij} \equiv m_{i}^2 - m_{j}^2$). Although intensively studied for over 25 years~\cite{Fukuda:1998mi, Fukuda:2002pe,Ahmad:2002jz, Eguchi:2002dm, Michael:2006rx, Ahn:2006zza, Abe:2011sj, Abe:2011fz, An:2012eh, Ahn:2012nd, Agafonova:2018auq}, key aspects of neutrino oscillations remain poorly constrained or entirely unknown. These include the determination of the neutrino mass ordering (whether \dmsq is positive or negative), the octant of \thatm (whether \thatm is less than or greater than $\pi/4$), and the existence of CP violation in the neutrino sector (quantified by \deltacp). 

Resolving these open questions and achieving precise measurements of the mixing parameters have broad implications for fundamental particle physics. The neutrino mass ordering (MO), i.e., whether the third mass eigenstate is heavier (normal MO) or lighter (inverted MO) than the others, is crucial for neutrino mass generation models, understanding supernova neutrino fluxes, and the search for neutrinoless double beta-decay (see~\cite{King:2014nza, Scholberg:2017czd, Dolinski:2019nrj} and references therein). The mixing angle \thatm governs the potential symmetry between the $\nu_\mu$ and $\nu_\tau$ components of $\nu_3$. 
The value of \deltacp is relevant to theories explaining the matter-antimatter asymmetry of the Universe via leptogenesis~\cite{Pascoli:2006ie, Branco:2006ce}.

The NuMI Off-axis $\nue$ Appearance (NOvA) experiment~\cite{Ayres:2007tu} is designed to study neutrino oscillations using Fermilab's NuMI beam~\cite{NuMI:Adamson2016} and two functionally identical detectors separated by a long baseline.
NOvA measures $P(\nu_\mu \rightarrow \nu_\mu)$ and $P(\nu_\mu \rightarrow \nu_e)$, along with their antineutrino counterparts, by comparing unoscillated neutrino interactions observed at the Near Detector (ND) with oscillated interactions at the Far Detector (FD).
The ND and FD are located \SI{1}{km} and \SI{810}{km}, respectively, from the neutrino beam target, with both detectors positioned \SI{14.6}{mrad} off axis, resulting in a narrow energy band peaked at \SI{1.8}{GeV}.

The Fermilab Main Injector accelerator delivers \SI{120}{GeV} protons that strike a \SI{1.2}{m} graphite target, producing a secondary hadron beam.
The hadrons are focused by magnetic horns and directed into a decay pipe, where they decay into neutrinos. Integrated over the 1--5~GeV energy range, the unoscillated beam is 94\% pure \numu in neutrino mode and 93\% pure \antinumu in antineutrino mode. The wrong-sign (WS) component, i.e., antineutrinos in neutrino mode and neutrinos in antineutrino mode, is 5\% (6\%) contamination in the neutrino (antineutrino) mode. The remaining flux is less than 1\% intrinsic \nue and \nuebar for both modes.

Both detectors are low-Z tracking calorimeters composed of PVC cells with a cross-sectional area of 6.6\timess \SI{3.9}{cm^2}, corresponding to a depth of 0.18 radiation lengths and, in the transverse direction, 0.4 Moli\`ere  radii~\cite{Talaga:2016rlq} with lengths of \SI{15.5}{m} (\SI{3.9}{m}) in the FD (ND).
These cells are assembled into planes alternating between vertical and horizontal orientations, enabling 3D event reconstruction.
Cells are filled with an organic scintillator (62\% of total detector mass) based on mineral oil with 5\% admixture of pseudocumene~\cite{Mufson:2015kga}. Scintillation light, along with a smaller amount of Cherenkov light, is collected by wavelength-shifting fibers looping through each cell with both fiber ends connected to a single pixel on an avalanche photodiode. 
The ND, located \SI{100}{m} underground, consists of 192 planes and features an additional downstream muon catcher composed of alternating steel and scintillator layers for a total active mass of \SI{193}{t}.
The \SI{14}{kt} FD consists of 896 planes. Located on the surface with an overburden of \SI{3.6}{m} water equivalent, the FD experiences a cosmic-ray flux of 130 kHz which constitutes the primary background. 
Data are recorded in a window centered around the 10\,$\mu$s NuMI beam spill~\cite{Norman:2012zza}.
In addition, cosmic-ray data are recorded for use in calibration and background estimation.   

The results presented in this Letter use data collected from February 6, 2014 to March 13, 2024, corresponding to a total FD neutrino (antineutrino) beam exposure of 26.61~$\times 10^{20}$ (12.50~$\times 10^{20}$) protons-on-target (POT) over \SI{885.53}{s} (\SI{322.59}{s}) of beam-on time. This is equivalent to a 96\% increase in the neutrino-mode beam exposure relative to previous NOvA results~\cite{NOvA:2021nfi, NOvA:2023iam}, while the antineutrino-mode exposure remains unchanged.

This analysis builds on previous NOvA work~\cite{NOvA:2021nfi}, incorporating updated simulation, improved systematic uncertainty treatments, and reoptimized energy reconstruction and event selection.

The neutrino flux is predicted at the ND and FD using \geant(v4.10)~\cite{Agostinelli:2002hh, Geant:2017ats}, which simulates particle production and transport through the NuMI beamline components,
and reweighted using the Package to Predict the FluX(\ppfx)~\cite{Aliaga:2016oaz}
to incorporate external constraints from hadron production measurements~\cite{Paley:2014rpb,Alt:2006fr,Abgrall:2011ae,Barton:1982dg,Seun:2007zz,Tinti:2010zz,Lebedev:2007zz,Baatar:2012fua,Skubic:1978fi,Denisov:1973zv,Carroll:1978hc,Abe:2012av,Gaisser:1975et,Cronin:1957zz,Allaby:1969de,Longo:1962zz,Bobchenko:1979hp,Fedorov:1977an,Abrams:1969jm}.
Neutrino interactions are simulated using a custom model configuration of \genie 3.0.6~\cite{Andreopoulos:2009rq,Andreopoulos:2015wxa} tuned to NOvA ND as well as external data.
This tuning creates a NOvA-specific version of the \valencia{} MEC model~\cite{PhysRevC.83.045501,Gran:2013kda}, which describes charged-current (CC) neutrino scattering from correlated nucleon pairs via meson exchange. 
The tune also uses pion--nucleus scattering data~\cite{Allardyce:1973ce,Saunders:1996ic,Meirav:1988pn,Levenson:1983xu,Ashery:1981tq,Ashery:1984ne,PinzonGuerra:2016uae} to improve the simulation of final-state interactions modeled with the \genie hN intranuclear cascade model~\cite{Andreopoulos:2015wxa}.
Outgoing final-state particles from neutrino interactions are then propagated through the detectors using \geant and a custom readout simulation~\cite{Aurisano:2015oxj}. 
 
\begin{figure*}[htb]
\makebox[\linewidth][c]{%
    \subfloat[\label{fig:numu_nue_spec-a}]{%
    \includegraphics[width=0.32\linewidth]{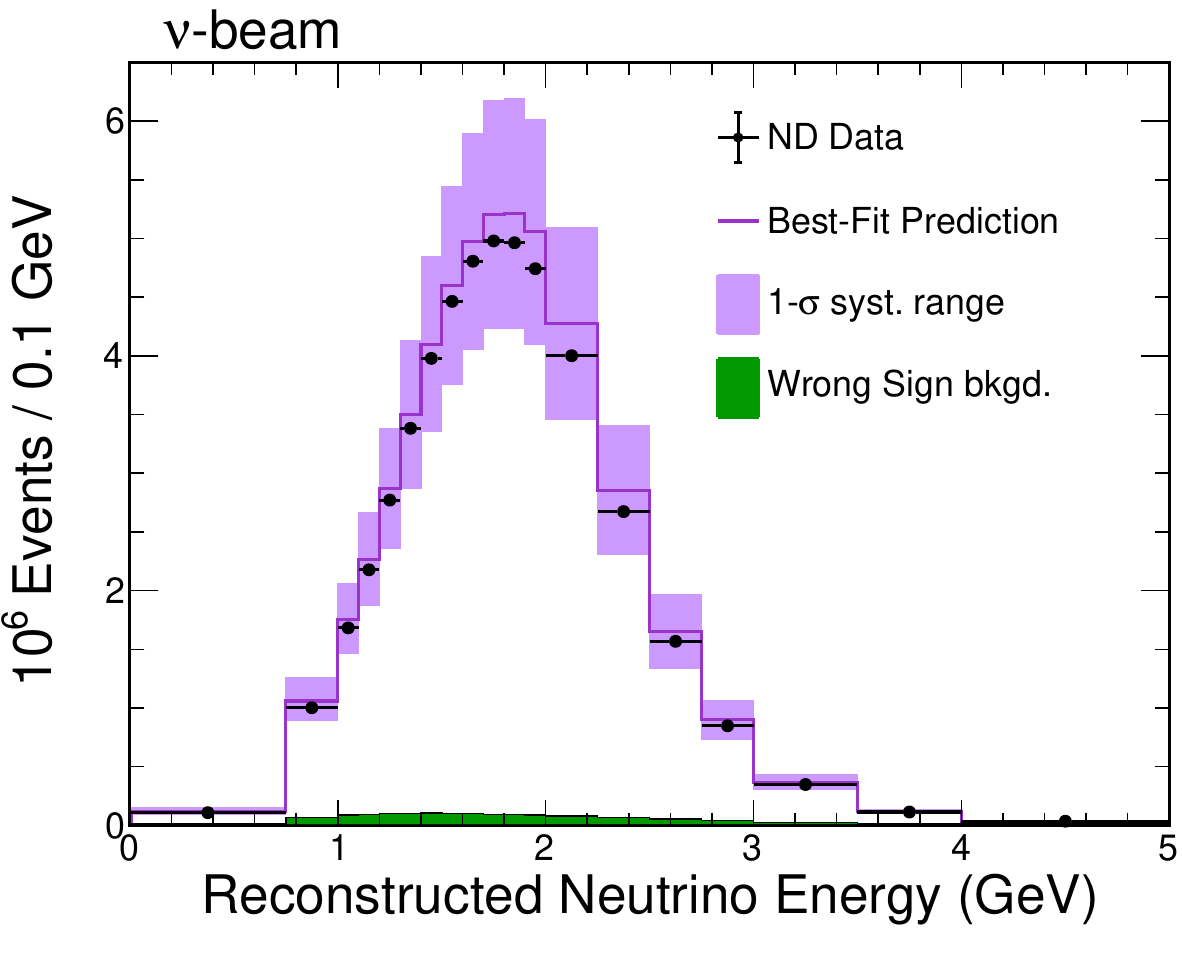}%
    }
    \hfill
    \subfloat[\label{fig:numu_nue_spec-b}]{%
    \includegraphics[width=0.32\linewidth]{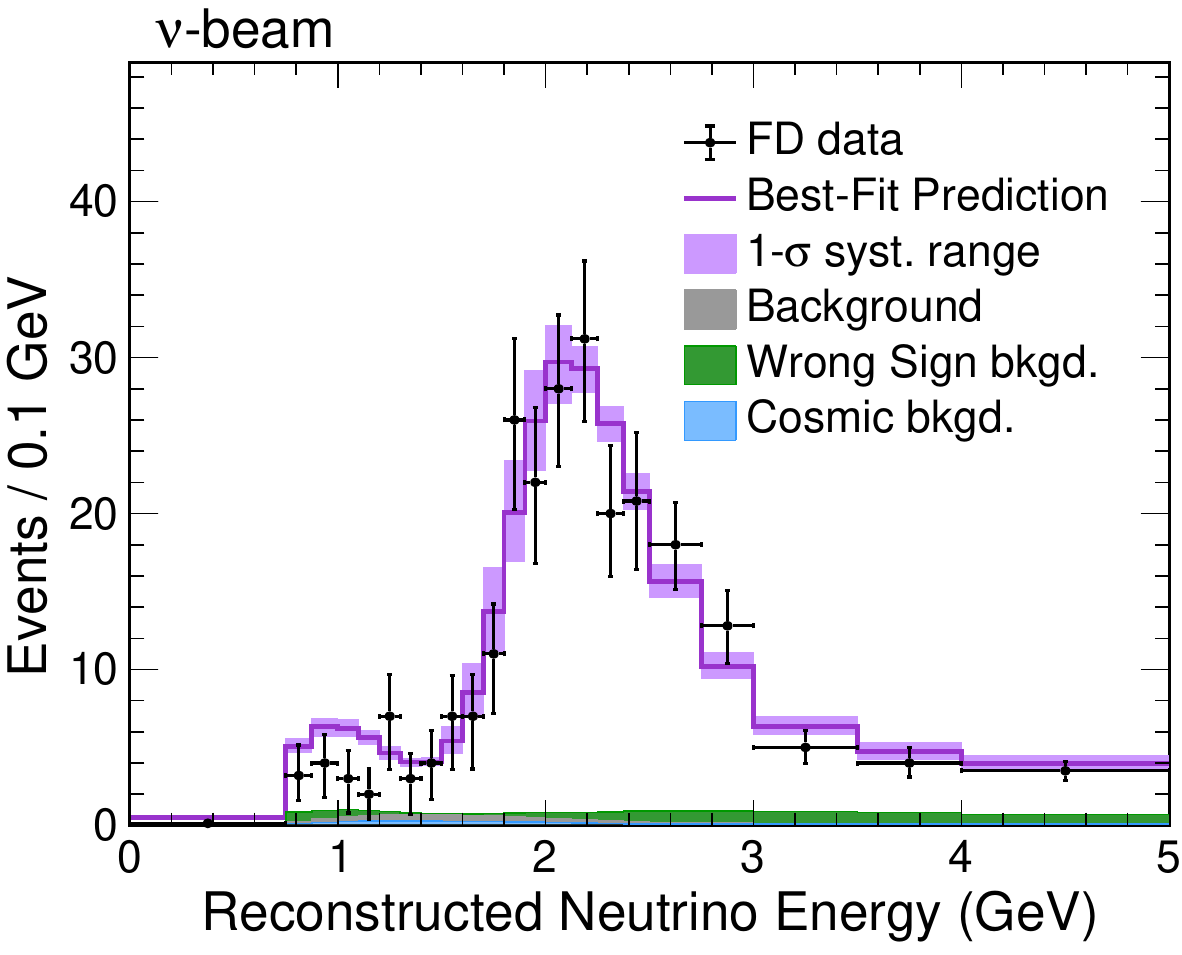}%
    }
    \hfill
    \subfloat[\label{fig:numu_nue_spec-c}]{%
    \includegraphics[width=0.32\linewidth]{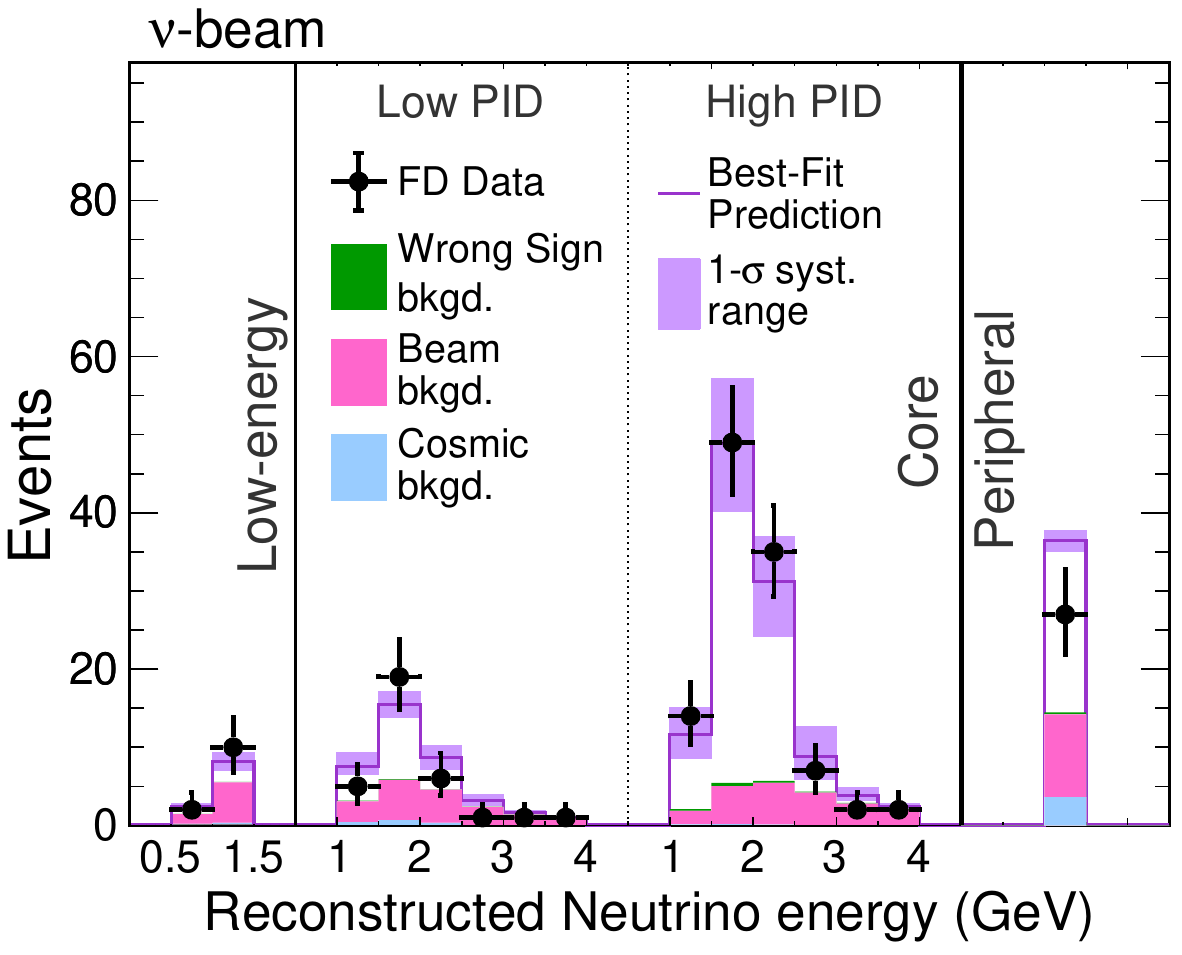}%
    }%
}
  \caption{Observed and predicted energy spectra, for $\nu_\mu$ CC selected events in the ND~(a) and FD~(b), and for $\nu_e$ CC selected events in the FD~(c). In the $\nu_\mu$ spectra all subsamples have been combined, while in the \nue spectra the four selections (Low-energy, low PID, high PID and peripheral) are shown separated. The best-fit prediction is extracted from the frequentist fit to the data with the Daya Bay 1D constraint on $\sin^2{2\theta_{13}}$~\cite{DayaBay:2022orm}. The predicted FD spectra have smaller systematic errors than the ND spectrum as a result of the extrapolation procedure.
  The corresponding antineutrino spectra and quartile-separated samples are shown in the supplemental material \cite{supplemental}.
  }
\label{fig:numu_nue_spec}
\end{figure*}

Energy distributions of neutrino candidates are built from reconstructed interactions in both detectors, and for both beam modes, in data and simulation. Detector cells with energy deposits above threshold are grouped into events and subsequently classified as \nue CC, \numu CC, neutral-current (NC), or cosmic candidates. This identification combines classical algorithms and deep-learning techniques such as a convolutional neural network, following the procedure established in the previous analysis~\cite{NOvA:2021nfi}. 
The \nue and \numu candidate samples include both beam-intrinsic and oscillated neutrinos in the FD, as well as WS neutrinos in each beam mode. They also contain misidentified events, primarily from NC interactions, cosmic rays, and other reconstruction failures. 
For brevity, \numuParen is used to indicate the dominant component in the neutrino (antineutrino) beam mode, although both modes contain a mix of \numu and \numubar.

The FD \numuParen samples primarily probe the atmospheric oscillation parameters \thatm and $|\dmsq|$.
The sensitivity from these samples depends on resolving the position and depth of the oscillation minimum (the ``dip'') in the reconstructed muon-neutrino energy spectrum, driven by $\accentset{(-)}{\nu}_\mu \rightarrow \accentset{(-)}{\nu}_\tau$ oscillations. Accurate estimation of the neutrino energy and careful choice of binning are critical to this measurement. 
This structure is visible in Fig.~\ref{fig:numu_nue_spec-b}, which shows the energy distribution for neutrino beam mode. 
The neutrino energy is estimated as the sum of the muon energy, determined from track length, and hadronic energy, estimated from calorimetry. The mean energy resolution is 6--13\% for muon neutrinos and 5--11\% for muon antineutrinos, where the variation depends on the relative contribution of the hadronic energy component.
To improve sensitivity to the dip structure, control systematic uncertainties, and account for acceptance differences between detectors, we use finer energy bins near the dip and divide the samples into bins of hadronic energy fraction and muon transverse momentum ($p_T$) \cite{NOvA:2021nfi}.

The FD \nueParen samples  primarily probe the MO, $\deltacp$ value, and \thatm octant. Electron-neutrino energy is estimated using a quadratic function of the hadronic and electromagnetic shower calorimetric energies. The overall energy resolution is approximately $11\%$ for electron neutrinos and $9\%$ for electron antineutrinos. 
The \nue appearance measurement is sensitive to separating signal from background due to a higher relative background contribution than in the \numu samples.
As a result, in addition to binning in reconstructed energy, two sub-samples of differing purities are constructed by grouping candidates into higher confidence (High PID) and lower confidence (Low PID) bins, where PID denotes the particle identification score.
These core bins are dominated by \nue CC events, arising either from beam-intrinsic \nue or from oscillated \numu.

The core bins are supplemented by the Peripheral and, in neutrino beam mode, Low-energy samples. As in the previous analysis~\cite{NOvA:2021nfi}, the Peripheral selection adds high-confidence \nue events with vertices outside the fiducial volume that may not deposit all their energy in the detector. 
This sample is treated as a single bin per beam mode rather than an energy distribution. 
Including these events increases the overall selection efficiency, which in turn provides a small improvement in sensitivity for certain combinations of $\deltacp$ and mass ordering, despite the higher cosmic background expected from their positions in the detector.
The Low-energy sample, introduced in this analysis, covers the 0.5 to 1.5~GeV energy range, a previously underexplored region with significant sensitivity to the MO, now accessible due to the cumulative increase in neutrino beam exposure. 
Although statistics in this region remain limited, they are sufficient to improve the oscillation constraints. 
The sample is expected to suffer from high backgrounds, primarily because  Low-energy $\nu_e$ events are difficult to distinguish from NC interactions. To recover these events, candidates that fail the core selection are passed through a boosted decision tree optimized to identify events with low-energy electromagnetic showers from primary electrons. Current statistics in antineutrino mode are insufficient to define a similar selection.

Similar reconstruction and selection techniques are applied at the ND to construct candidate samples. In the neutrino beam mode, the \numu ND candidate distributions are used to constrain both the surviving \numu and appearing \nue FD signal events while the \nue ND candidates are used to constrain the various FD beam backgrounds in the \nue sample. An analogous process is used for antineutrino beam samples. This data-driven method for building FD predictions prior to fitting the FD data to infer oscillation parameters is referred to as extrapolation and remains unchanged from the previous analysis~\cite{NOvA:2021nfi}. This procedure is essential for mitigating the impact of systematic uncertainties at the FD through the use of high-statistics ND data.

Systematic uncertainties are evaluated by varying model parameters in the simulation. The resulting shifted samples are processed through the same reconstruction, selection, and extrapolation procedures to assess their impact on the analysis. In this work, systematic uncertainties are grouped into categories including: neutrino flux, neutrino interaction cross sections, detector simulation, calibrations, lepton reconstruction, uncorrelated effects between the ND and FD, and neutron propagation in the detector.  Most uncertainty estimates are unchanged from the previous analysis~\cite{NOvA:2021nfi}, although several new sources are now included.

In the cross-section category, new uncertainties are introduced to address potential mismodeling of pion production in the transition region between resonant (RES) and deep inelastic scattering (DIS), which is particularly challenging to model accurately~\cite{NuSTEC:2017hzk}. \genie's implementation assumes, as an ansatz, a fixed relationship between charged and neutral final states in this region. It also does not include uncertainties on the relative production rates of $\Delta$ and higher resonances. Therefore, additional uncertainties are included to account for possible variations in these quantities consistent with the current data~\cite{MartinezCasales:2023bkf,Dolce:2023ije}.
For the detector model, the \geantrwgt package~\cite{Calcutt:2021zck} is used to address uncertainties in the inelastic scattering of protons and pions produced during both primary neutrino interactions and subsequent secondary interactions \cite{Sweeney:2023mzt}; these uncertainties are incorporated using principal component analysis \cite{Jargowsky:2024mwc}.
In the neutron model category, a supplementary model, \menate~\cite{roeder2008development,KOHLEY201259}, is introduced to evaluate neutron detection uncertainties based on its differences with the default \geant model \cite{Rabelhofer:2023cyp}. 
\menate provides improved modeling of neutron interactions in the 20--200~MeV range, where \geant tends to underestimate energy deposition in scintillator detectors.
Finally, following the update to NOvA’s custom light model, two sets of systematic uncertainties are applied: a $\pm$5\% variation in the overall light yield and a $\pm$6.2\% scaling of the Cherenkov light-collection efficiency.

The extrapolation method reduces the impact of systematic uncertainties that are correlated between the detectors, particularly those related to flux and cross section. To provide a scale of this improvement, systematic uncertainties on the predicted total FD event rates drop from $12\textrm{--}18$\% to $3\textrm{--}6.5$\%, depending on the sample.
Among the systematic uncertainties, those related to detector calibration contribute most to the uncertainties in \dmsq and \snsq, while cross-section systematics, especially those associated with the uncertainty in \nue/\numu and \nuebar/\numubar cross section, primarily affect the \deltacp measurement. 
Table~\ref{table:systs} summarizes the statistical and systematic contributions to the oscillation parameter uncertainties, showing that statistical errors are the dominant source of uncertainty for all measured parameters.

\begin{table}[htb]
  \centering
  \caption{Systematic and statistical uncertainties on the oscillation parameters $\Delta m^2_{32}$, $\sin^2 \theta_{23}$, and $\delta_{\rm CP}$ are evaluated at the best-fit point extracted from a frequentist fit with the Daya Bay 1D constraint on $\sin^2{2\theta_{13}}$~\cite{DayaBay:2022orm}. These predicted $\pm 1 \sigma$ ranges are given as absolute error on the best-fit values. Asymmetric errors on $\delta_{\rm CP}$ and $\sin^2 \theta_{23}$ arise from limited sensitivity and degeneracy in parameters, leading to several highly probable regions that vary with the best-fit values.} 
  \label{table:systs}
  \begin{tabular}{l@{\hspace{0.3cm}}c@{\hspace{0.3cm}}c@{\hspace{0.3cm}}c}
       \hline \hline
  \rule{0pt}{9pt} & $|\dmsq|$ & \snsq & $\deltacp $ \\
  Source & $(\times10^{-5} ~\text{eV}^2)$ & $(\times10^{-3} )$ & $(\times10^{-2} ~\pi )$ \\
  \hline
  Beam flux   & +0.06/-0.12      & +0.40/-0.75    & +0.16/-1.00  \\
  Calibration & +1.40/-1.80      & +3.20/-16.0    & +1.70/-17.0 \\
  Detector model & +0.17/-0.27   & +0.36/-3.90     & +0.37/-3.30  \\
  Lepton Reco.   &  +0.90/-1.40  &  +2.70/-4.50    & +0.61/-2.90 \\
  ND -- FD Uncor. &  +0.25/-0.34 & +2.40/-2.50    & +0.79/-4.00 \\
  Cross Sections  & +0.65/-1.10  & +3.10/-5.10    & +2.00/-9.70 \\
  Neutron model   & +0.25/-0.39  & +2.70/-0.49    & +0.43/-1.30  \\
  \hline
  Systematic Unc.  & +1.70/-2.30 & +6.60/-18.0   & +2.90/-21.0 \\
  Statistical Unc. & +3.20/-4.40 & +22.0/-85.0  & +6.60/-74.0 \\
  \hline \hline
  \end{tabular}
\end{table}

Oscillation parameters are inferred using a Poisson likelihood with a log-likelihood difference test statistic to compare the data and the corrected simulated FD spectra across various oscillation and systematic hypotheses. 
The analysis constrains $\sin^2 \theta_{23}$, $\sin^2 2\theta_{13}$, $\Delta m^2_{32}$, and $\delta_{\rm CP}$, with both hypotheses of neutrino MO tested against the data. 
Because the sensitivity to $\theta_{13}$ is limited when fitting multiple oscillation parameters simultaneously, external input from the Daya Bay experiment~\cite{DayaBay:2022orm} is incorporated to improve precision.
Three treatments of $\theta_{13}$ are considered: no constraint (using only NOvA data), the Daya Bay $\sin^2 2\theta_{13}$ constraint (1D Daya Bay), and the two-dimensional Daya Bay constraint on $\sin^2 2\theta_{13}$ and $\Delta m^2_{32}$ (2D Daya Bay). All other oscillation parameters are fixed to their values from PDG 2023~\cite{ParticleDataGroup:2022pth}. 

In this Letter, results are extracted using two complementary analyses -- a Bayesian Markov Chain Monte Carlo (MCMC) method to compute posterior probability density distributions, and a frequentist analysis that yields consistent results.
The Bayesian fits use uniform priors in the oscillation parameters as plotted (e.g., flat in $\sin^2\theta_{23}$ and $\delta_{\textrm{CP}}$) and Gaussian priors for the systematic parameters, as detailed in~\cite{NOvA:2023iam}.
The results from these fits can be marginalized (in the Bayesian framework) or profiled (in the frequentist framework), either simultaneously over both MO (non-conditional) or separately for each MO (conditional). Unless stated otherwise, the results reported in the main text are conditional on the MO and incorporate the 1D Daya Bay constraint on $\theta_{13}$. A comprehensive set of results, including different marginalizations over MO and variations in constraints from Daya Bay, is provided in the supplemental materials~\cite{supplemental}.
  
With the increased statistics and updated analysis, NOvA recorded 181\,$\nu_e$ and 384\,$\nu_\mu$ candidates in neutrino mode, and 32\,$\bar{\nu}_e$ and 106\,$\bar{\nu}_\mu$ candidates in antineutrino mode at the FD.
The FD spectra resulting from the fitting procedure are shown in Fig.~\ref{fig:numu_nue_spec} for the neutrino beam, and Table~\ref{table_data_and_mc} presents the observed neutrino counts alongside the Monte-Carlo predictions. These predicted values, separated into signal and background sub-components, are derived from the frequentist best-fit oscillation and systematic parameters.
These fits yield $\dmsq=2.441^{+0.035}_{-0.031}\times 10^{-3}$~eV$^2$ in the normal MO and $\dmsq=-2.481^{+0.034}_{-0.034} \times 10^{-3}$~eV$^2$ in the inverted MO. The inverted MO is disfavored at 1.4\,$\sigma$ confidence level. The best-fit point for $\delta_{\textrm{CP}}$ is $0.87^{+0.30}_{-0.90}$~$\pi$ and $\sin^2\theta_{23} = 0.55^{+0.02}_{-0.06}$. The Feldman--Cousins approach~\cite{Feldman:1997qc, NOvA:2022wnj} is used to determine frequentist confidence intervals for the oscillation parameters.

\begin{table}[htbp] 
  \centering
  \caption{Observed and predicted numbers of $\nu_\mu$, $\nu_e$ and Low-energy $\nu_e$ events in the neutrino beam, and $\bar{\nu}_\mu$ and $\bar{\nu}_e$ events in the antineutrino beam. The low PID, high PID, and Peripheral samples have been combined in the \nueParen columns, while the Low-energy sample is shown separately due to its novel status in this analysis. WS is considered signal in the \numu and \numubar samples, but background in the \nue and \nuebar samples. We denote as Others the sum of all remaining oscillation channels that are not listed explicitly in this table. The best-fit point for the prediction is extracted from a frequentist fit to the data with the Daya Bay 1D constraint on $\sin^2{2\theta_{13}}$~\cite{DayaBay:2022orm}.}
  \label{table_data_and_mc}
  \begin{tabular}{lccccc}
  \toprule \toprule
  \multirow{3}{*}{} & \multicolumn{3}{c}{Neutrino} & \multicolumn{2}{c}{Antineutrino} \\
   & \multicolumn{3}{c}{Beam} & \multicolumn{2}{c}{Beam} \\
  \cmidrule(lr){2-4} \cmidrule(lr){5-6}
   \multirow{2}{*}{Sample} & \multirow{2}{*}{$\nu_\mu$} & \multirow{2}{*}{$\nu_e$} &  Low- & \multirow{2}{*}{$\bar{\nu}_\mu$} & \multirow{2}{*}{$\bar{\nu}_e$} \\
   &  &  &  energy &  &  \\
  \midrule
  \numu$\rightarrow$\numu       & 372.3 & 4.3   & 0.3 & 24.4 & 0.2 \\
  \numubar$\rightarrow$\numubar & 24.5  & 0.1   & 0.0 & 71.5 & 0.2 \\
  \numu$\rightarrow$\nue        & 0.4   & 125.3 & 3.4 & 0.0  & 2.1 \\
  \numubar$\rightarrow$\nuebar  & 0.0   & 1.8   & 0.1 & 0.0  & 18.9 \\
  Beam $\nue+\nuebar$  & 0.1   & 26.1  & 0.8 & 0.0  & 6.5 \\
  NC                            & 5.5   & 16.8  & 5.3 & 0.8  & 2.0 \\
  Cosmic                        & 4.4   & 5.5   & 0.5 & 0.7  & 1.1 \\
  Others                        & 1.5   & 0.8   & 0.1 & 0.2  & 0.1 \\
  \midrule
  \rule[-4pt]{0pt}{13pt}Signal & $397.6^{+26.8}_{-24.2}$ & $125.3^{+5.1}_{-5.1}$ & $3.4^{+1.0}_{-0.8}$ & $96.0^{+5.2}_{-5.7}$ & $18.9^{+0.8}_{-0.9}$ \\
  \rule[-4pt]{0pt}{13pt}Background & $11.0^{+1.3}_{-1.2}$ & $55.4^{+2.6}_{-1.7}$ & $7.1^{+1.3}_{-1.1}$ & $1.7^{+0.2}_{-0.2}$ & $12.2^{+0.5}_{-0.7}$ \\
  \midrule
  \rule{0pt}{9pt}Predicted & 408.6 & 180.7 & 10.5 & 97.7 & 31.1 \\
  Observed & 384 & 169 & 12 & 106 & 32 \\
  \bottomrule \bottomrule
  \end{tabular}
\end{table}

\begin{figure}[htb]
  \centering
  \includegraphics[width=\linewidth]{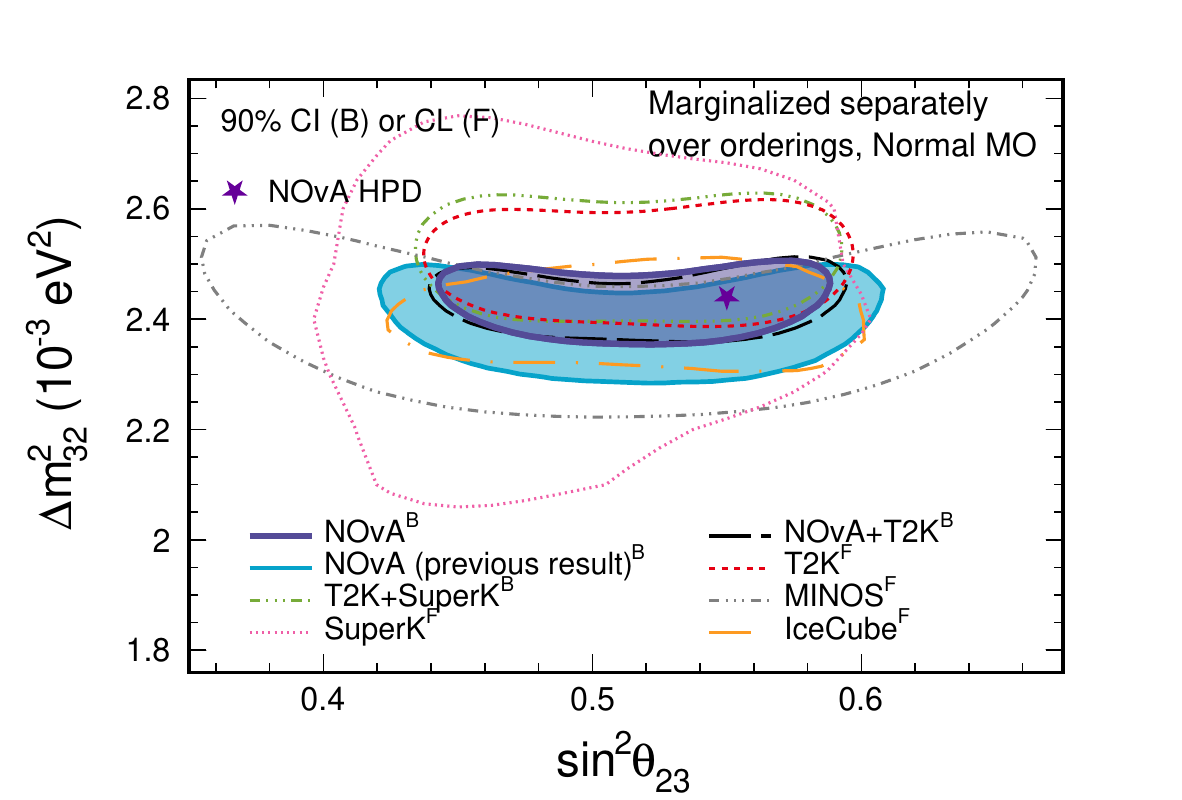}
  \caption{Comparisons of the 90\% intervals for $\Delta m^2_{32}$ --
           $\sin^2\theta_{23}$ in the normal MO with NOvA 2020
           results and superimposed contours from other experiments~\cite{T2K:2024wfn,  IceCube:2024xjj, t2k_collaboration_2022_6908532, MINOS:2020llm, Super-Kamiokande:2023ahc}, including the 2024 joint NOvA-T2K analysis~\cite{novat2k_nature2025}.The NOvA results are with the 1D Daya Bay constraint on $\theta_{13}$ applied~\cite{DayaBay:2022orm}. Contours labeled B are from Bayesian analyses, while those labeled F are from frequentist analyses, used when Bayesian results were not available.}
  \label{fig:with_friends}
\end{figure}

The Bayesian MCMC fits report
 $\dmsq=2.431^{+0.036}_{-0.034} \times 10^{-3}$~eV$^2$ in the normal MO or $\dmsq=-2.479^{+0.036}_{-0.036}$ $\times 10^{-3}$~eV$^2$ in the inverted MO. 
 The highest posterior density (HPD) for $\delta_{\textrm{CP}}$ is $0.93^{+0.21}_{-0.89}$~$\pi$ in the normal MO.
For both MO, the data favor regions of parameter space close to maximal mixing, $\sin^2\theta_{23} = 0.55^{+0.02}_{-0.06}$, with no significant preference for octant. Fig.~\ref{fig:with_friends} shows the 90\% credible regions in the $\dmsq$--$\sin^2\theta_{23}$ plane extracted from this analysis compared to other experiments and previous NOvA results. Good agreement is seen across all the experiments. 
The goodness of fit is estimated using posterior-predictive $p$-values (PPP), where sampled models that describe the data well will have values close to 0.5~\cite{Gelman:1996}. With the 1D Daya Bay constraint on $\theta_{13}$, the total PPP across all data samples is 0.48, indicating good agreement between the data and the model. The PPP does not change significantly when either no external constraint on $\theta_{13}$ is used or when the two-dimensional $\sin^2 2\theta_{13}$–$\Delta m^2_{32}$ constraint from Daya Bay is applied~\cite{supplemental}.

\begin{figure}[htb]
\includegraphics[width=0.90\linewidth]{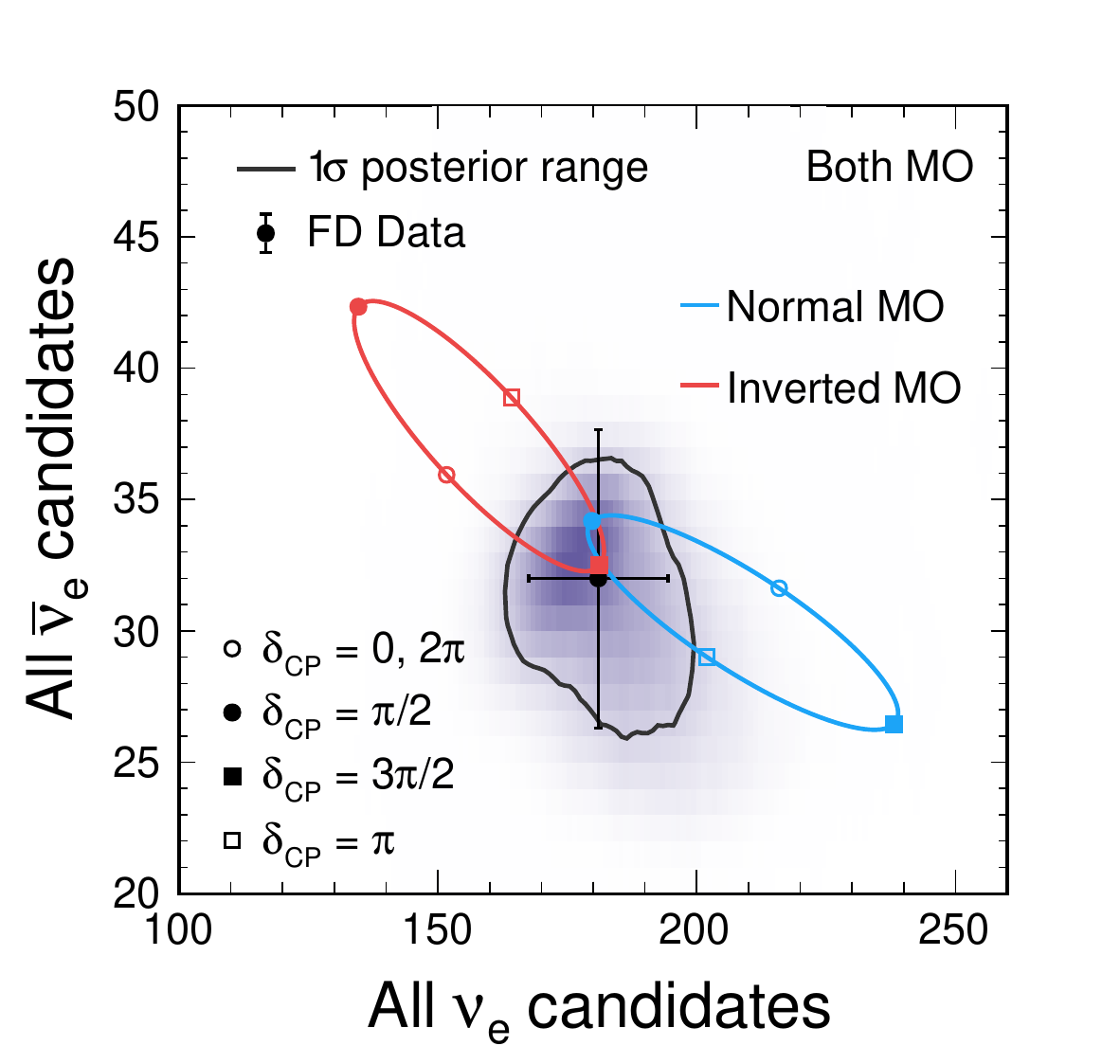}
  \caption{Bi-event plot showing the posterior probability for the predicted
           number of $\nu_e$ and $\bar{\nu}_e$ events in purple, with the
           1\,$\sigma$ credible interval and the FD data point marked. The two ovals, red for the inverted MO and blue for the normal MO, show the prediction with all the parameters fixed at the NOvA best-fit, varying only $\delta_{\textrm{CP}}$, with four $\delta_{\textrm{CP}}$ points marked.}
\label{fig:bievent}
\end{figure}

The NOvA data lie in the degenerate region of parameter space which can be accommodated within either MO by varying the preferred values of $\delta_{\rm CP}$, as shown in Fig.~\ref{fig:bievent}. Bayes factors, representing ratios of posterior probabilities between two hypotheses, are used here to represent how much more probable one model is over another. Bayes factors for the preference of CP violation over CP conservation were extracted using the Savage-Dickey method~\cite{Dickey:1971, Mulder:2020} based on the Jarlskog invariant~\footnote{The magnitude of CP violation is dependent on all the neutrino mixing angles, rather than $\delta_{\textrm{CP}}$ alone, and is dependent on the parametrization of the PMNS matrix. Jarlskog invariant is a rephase-invariant measure, and therefore more useful in determining non-conservation of the CP symmetry.}, with Jarlskog invariant of 0 representing CP conservation. The Bayes factors for CP violation are 1.1 (containing 51\% of the posterior probability) in the normal MO and 4.3 (81\%) in the inverted MO, and remain compatible with both CP conservation and CP violation.

\begin{figure}[htb]
\includegraphics[width=\linewidth]{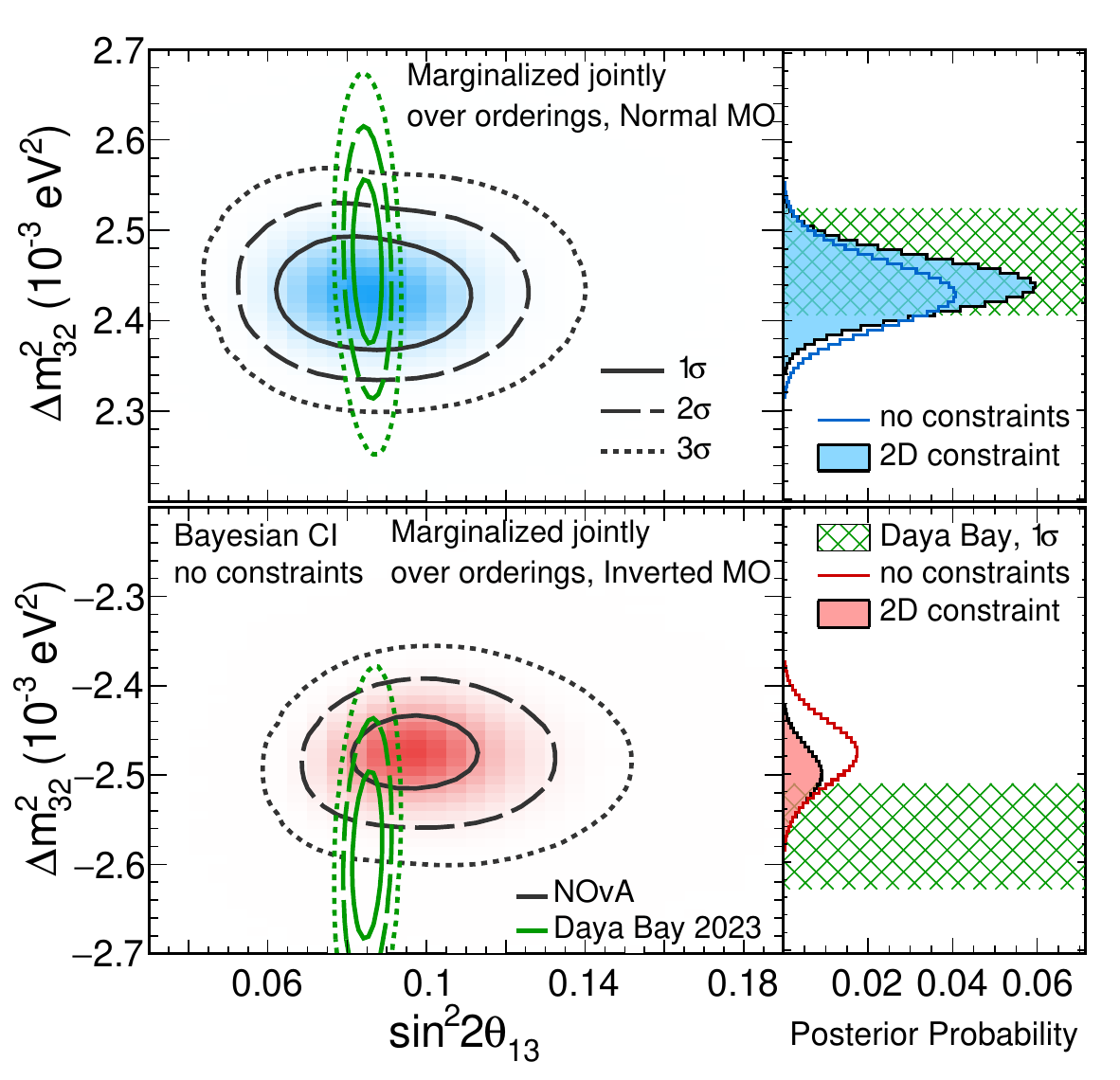}
  \caption{A comparison of the NOvA and Daya Bay constraints in the
           $\sin^2 2 \theta_{13}$ -- $\dmsq$ plane under the assumption of
           normal (top) or inverted (bottom) MO. The 2D binned NOvA
           posterior shown in color in the left panels is without any external constraints on $\theta_{13}$.
           Panels on the right illustrate the effect of applying a 2D $\sin^2\theta_{13}\textrm{--}\Delta m^2_{32}$
constraint from Daya Bay~\cite{DayaBay:2022orm} on the MO preference.}
\label{fig:nova_dayabay_2d}
\end{figure}

This analysis reports the most precise single-experiment measurements of \dmsq, achieving a fractional uncertainty of 1.5\%.  
This improved precision provides an additional lever for probing the neutrino MO~\cite{PhysRevD.72.013009,Parke:2024xre}. Reactor and long-baseline accelerator measurements of $\Delta m^2_{32}$ should agree only under the correct MO hypothesis. Fig.~\ref{fig:nova_dayabay_2d} shows a comparison of the NOvA and the 2D Daya Bay constraints in the $\sin^2 2 \theta_{13}\textrm{--}\dmsq$ plane under the two MO hypotheses, with somewhat better agreement in the normal MO. Without any external $\theta_{13}$ constraint, NOvA data show a slight preference for normal MO vs inverted MO, with a Bayes factor of 2.4 (70\% of posterior probability in favor of normal MO over inverted MO hypothesis). Applying the 1D $\sin^2 \theta_{13}$ constraint from Daya Bay~\cite{DayaBay:2022orm}, this preference increases to 3.3 (77\%). When a 2D $\sin^2\theta_{13}\textrm{--}\Delta m^2_{32}$
constraint from Daya Bay is applied, the preference strengthens further to 6.6 (87\%). The posterior probability distributions for each are shown in the supplemental materials~\cite{supplemental}, and the data release is available at~\cite{datarelease_nova2024}.

To conclude, NOvA has reported improved constraints on $\delta_{\rm CP}$, $\theta_{13}$, $\theta_{23}$ and $\Delta m^2_{32}$ with 10 years of data collected with neutrino and antineutrino beams using enhanced analysis techniques. 
Notably, NOvA now provides the most precise measurement of $\Delta m^2_{32}$ to date, with a precision of 1.5\%.
This enhanced precision in measuring $\Delta m^2_{32}$ enables the complementarity between accelerator and reactor experiments to be leveraged by incorporating a 2D $\sin^2\theta_{13}\textrm{--}\Delta m^2_{32}$ constraint from Daya Bay, strengthening the preference for the normal MO to 87\%.
Additional data from current and upcoming neutrino oscillation experiments will help resolve degeneracies, enabling a definitive determination of the MO and the potential discovery of CP violation in the lepton sector. High-precision measurements of mixing parameters across multiple oscillation channels will also provide a robust test of the three-flavor paradigm in neutrino physics.

\section*{Acknowledgements}

\begin{acknowledgments}
This document was prepared by the NOvA collaboration using the resources of the Fermi National Accelerator Laboratory (Fermilab), a U.S. Department of Energy, Office of Science, HEP User Facility. Fermilab is managed by Fermi Forward Discovery Group, LLC, acting under Contract No. 89243024CSC000002.  
This research used resources of the National Energy Research Scientific Computing Center, a DOE Office of Science User Facility supported by the Office of Science of the U.S. Department of Energy under Contract No. DE-AC02-05CH11231 using NERSC award HEP-ERCAP0028967.
This work was supported by the U.S. Department of Energy; the U.S. National Science Foundation; the Department of Science and Technology, India; the European Research Council; the MSMT CR, GA UK, Czech Republic; the RAS, the Ministry of Science and Higher Education, and RFBR, Russia; CNPq and FAPEG, Brazil; UKRI, STFC and the Royal Society, United Kingdom; and the State and University of Minnesota.  We are grateful for the contributions of the staffs of the University of Minnesota at the Ash River Laboratory, and of Fermilab. For the purpose of open access, the authors have applied a Creative Commons Attribution (CC BY) license to any Author Accepted Manuscript version arising.

\end{acknowledgments}

\bibliographystyle{apsrev4-1}
\bibliography{cites}

\clearpage

\clearpage
\setcounter{page}{1} 
\setcounter{secnumdepth}{3} 

 \widowpenalty=10000
 \clubpenalty=10000
\brokenpenalty=10000
\onecolumngrid
\renewcommand{\thepage}{S\arabic{page}} 
\renewcommand{\thesection}{}
\renewcommand{\thesubsection}{\Roman{subsection}}
\renewcommand{\thesubsubsection}{\thesubsection:\arabic{subsubsection}}

\renewcommand{\thefigure}{S\arabic{figure}}   
\renewcommand{\thetable}{S\arabic{table}}     
\renewcommand{\theequation}{S\arabic{equation}} 

\setcounter{section}{0}
\setcounter{subsection}{0}
\setcounter{figure}{0}
\setcounter{table}{0}
\setcounter{equation}{0}

\newcommand{\figwidth}{0.5\linewidth}
\newcommand{\largefigwidth}{0.8\linewidth}

\section*{Supplemental Material for: \\ \papertitle}

This supplemental material provides additional details supporting the main results presented in the Letter. Sec.~\ref{sec:nova_data} presents further NOvA data distributions used in the analysis. 
Sec.~\ref{sec:results} provides a detailed presentation of the Bayesian analysis, with Table~\ref{table_BF_and_LL} summarizing the highest posterior density points and $1\sigma$ intervals under various assumptions. 
Sec.~II:1 enumerates the goodness of fit using posterior-predictive $p$-values, split by FD data sample and reactor constraints. 
Sec.~II:2 provides one- and two-dimensional posterior probability distributions, while Sec.~II:3 compares results across several other experiments. Sec.~II:4 briefly discusses the MO and $\theta_{23}$ octant preferences derived from the analysis. 
Finally, Sec.~\ref{sec:freq_results} presents results obtained using an alternative frequentist statistical treatment of the dataset.

\subsection{NOvA Data} \label{sec:nova_data}

\begin{figure*}[htb]
\makebox[\linewidth][c]{%
    \subfloat[\label{fig:numubar_nuebar_spec-a}]{%
        \includegraphics[width=0.32\linewidth]{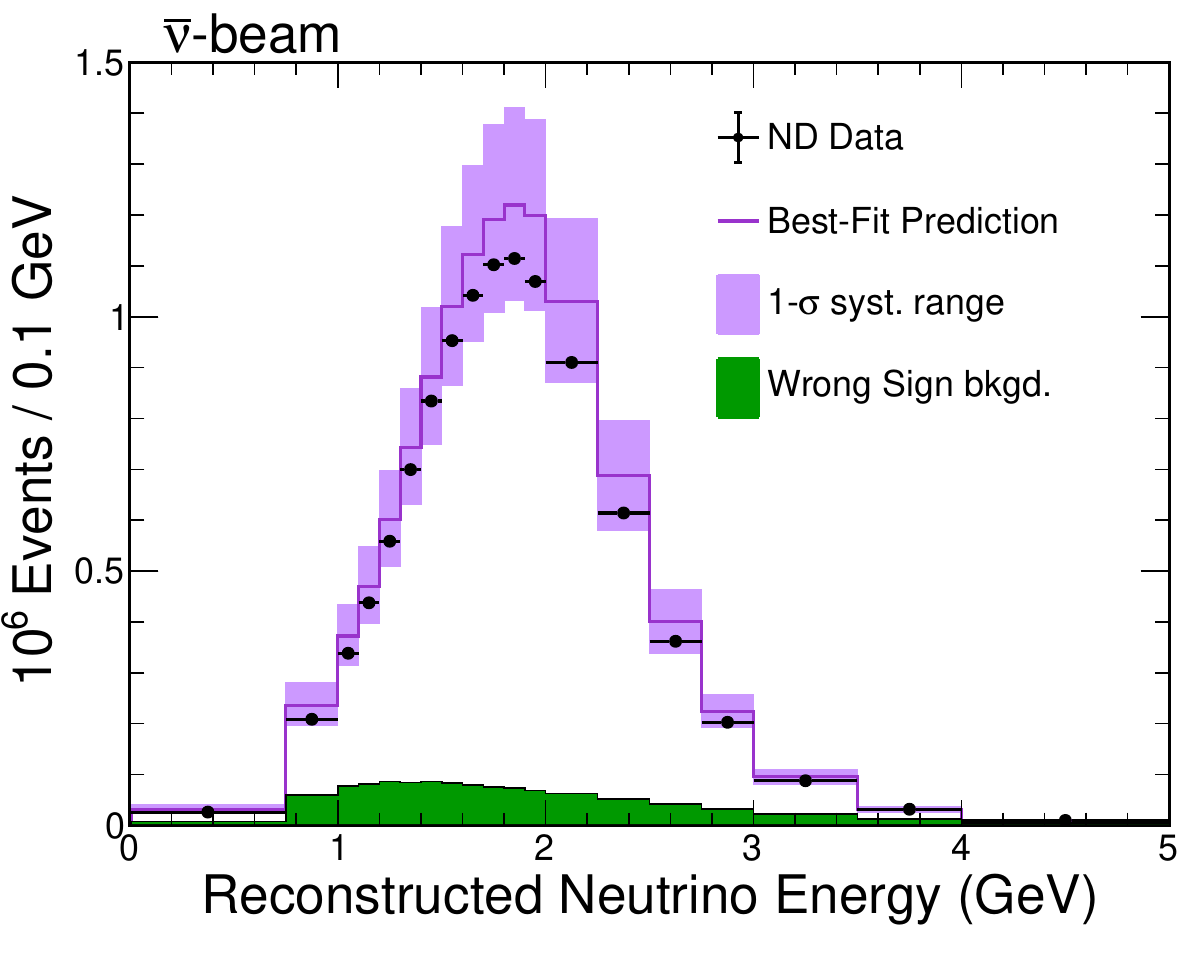}%
    }%
    \hfill
    \subfloat[\label{fig:numubar_nuebar_spec-b}]{%
        \includegraphics[width=0.32\linewidth]{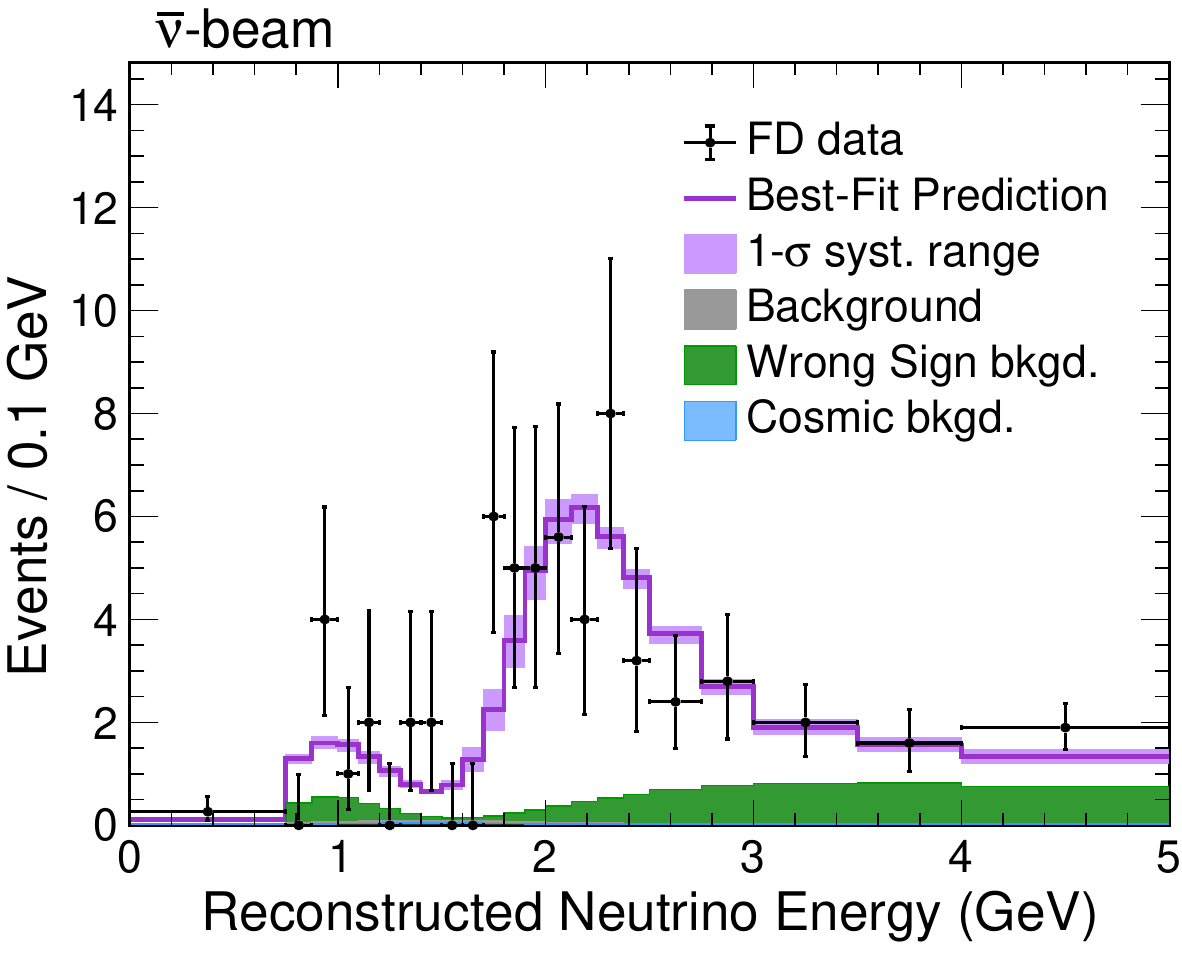}
    }%
    \hfill
    \subfloat[\label{fig:numubar_nuebar_spec-c}]{%
        \includegraphics[width=0.32\linewidth]{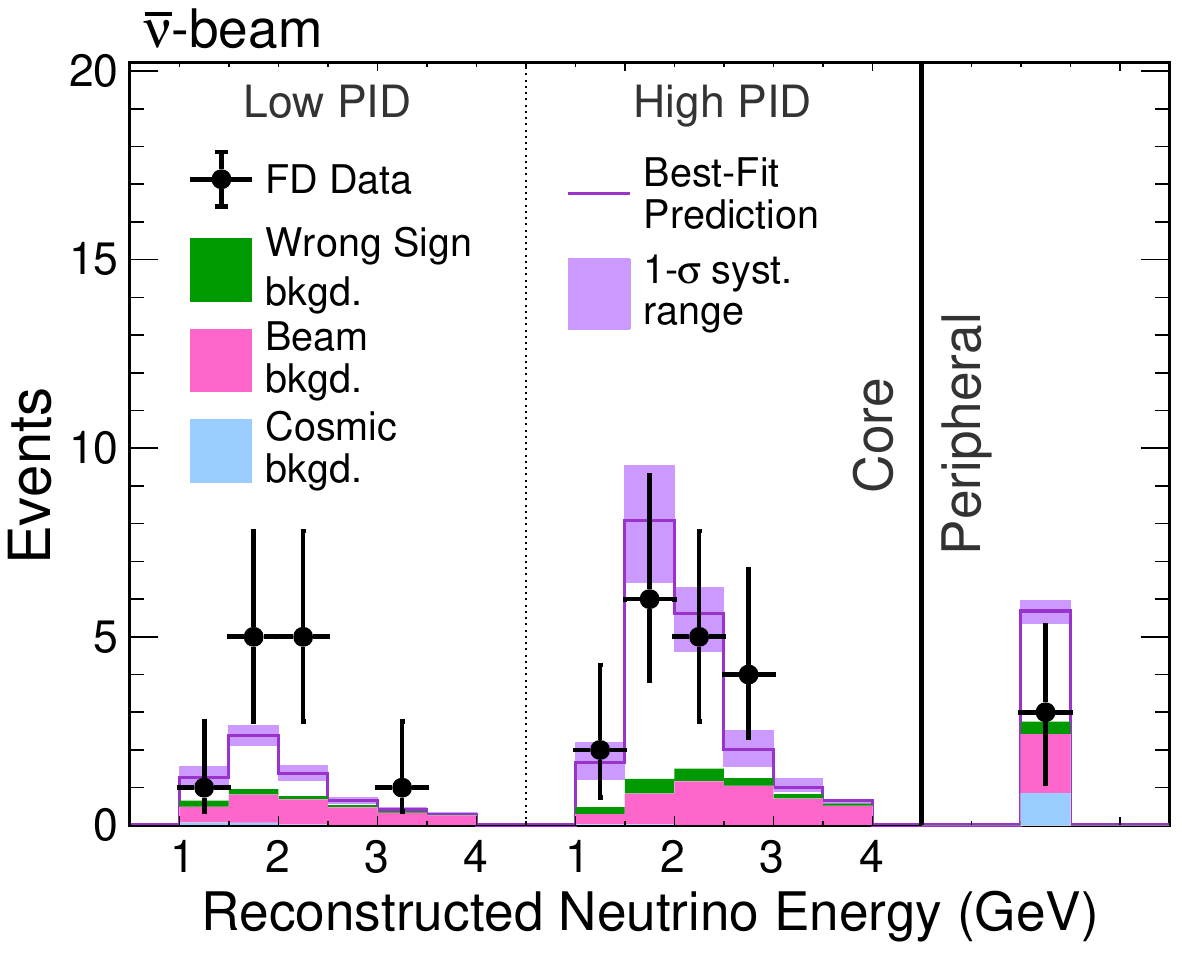}
    }%
}
    \caption{Observed and predicted energy spectra, for \numubar CC selected events in the ND (left) and FD (middle), and for \nuebar CC selected events in the FD (right). In the \numubar spectra all hadronic energy fraction quartiles and transverse momentum quantiles have been combined, while in the \nuebar spectra the three selections (low PID, high PID, and peripheral) are shown separated. The best-fit prediction is extracted from a frequentist fit to the data (best-fit values in Table~\ref{table_freq_BF}) with the Daya Bay 1D constraint on $\sin^2 2\theta_{13}$ \cite{DayaBay:2022orm}.}
\label{fig:numubar_nuebar_spec}
\end{figure*}

\begin{figure*}[htb]
\includegraphics[width=0.75\linewidth]{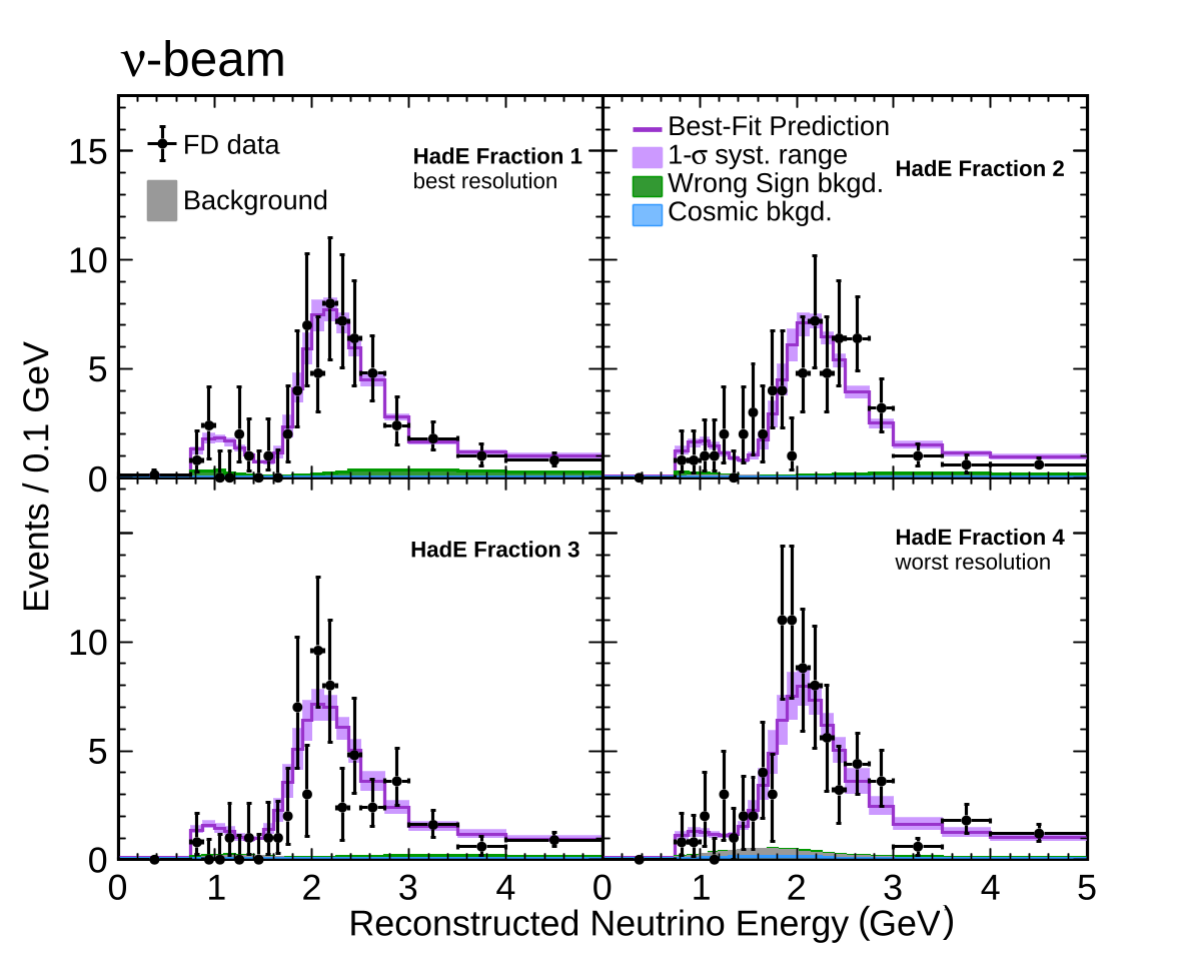}
\includegraphics[width=0.75\linewidth]{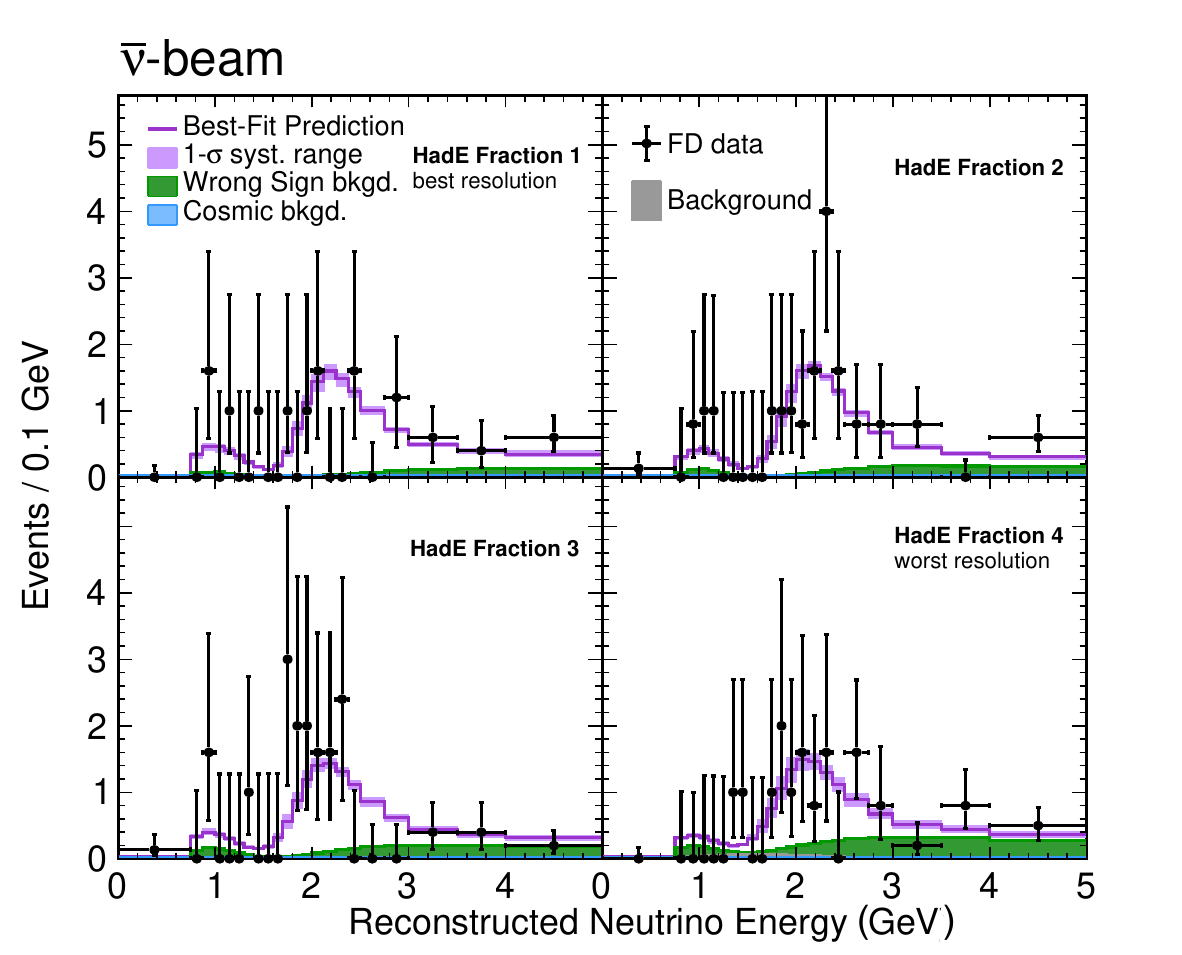}
  \caption{Observed and predicted energy spectra for \numu (left) and \numubar
  (right) CC selected events in the FD, split into four
  hadronic energy fractions but with the transverse momentum quantiles
  combined. The best-fit prediction is extracted from a frequentist fit to the data (best-fit values in Table~\ref{table_freq_BF}) with the Daya Bay 1D constraint on $\sin^2 2\theta_{13}$ \cite{DayaBay:2022orm}.}
\label{fig:fd_spec_byquartile}
\end{figure*}

The data collected by the NOvA experiment used for the analysis shown in
the associated Letter are described in the analysis section, with
Fig.~\ref{fig:numu_nue_spec} showing the observed and predicted energy
spectra in the neutrino beam mode. For completeness,
Fig.~\ref{fig:numubar_nuebar_spec} shows the observed and predicted energy
spectra in the antineutrino beam mode, and Fig.~\ref{fig:fd_spec_byquartile}
further splits the \numu and \numubar spectra into the four hadronic energy
fractions used in the analysis.

\FloatBarrier

\subsection{Bayesian Results} \label{sec:results}

A summary of the neutrino oscillation parameters measured in the associated Letter is shown in Table~\ref{table_BF_and_LL}. These numbers are obtained from the Bayesian MCMC sampling and represent the highest posterior density areas and 1\,$\sigma$ credible intervals. The constraints are shown for both Normal and Inverted MOs, and for each of the three treatments of the reactor constraint: no external constraints; one-dimensional reactor constraint on $\theta_{13}$ from Daya Bay; and the two-dimensional reactor constraint on $\theta_{13}$ and $\Delta m^2_{32}$ from Daya Bay \cite{DayaBay:2022orm}.

\begin{table*}[!htbp] 
  \centering
  \caption{Summary of the oscillation parameters measured by NOvA for various reactor constraint options. The indicated regions correspond to 1$\sigma$ intervals for the highest posterior density values. All other parameters are fixed at the PDG 2023 central values. The results for marginalizing over both MOs and for marginalizing separately over each MO are shown, though the HPD is always in the Normal Ordering under non-conditional marginalization (we denote it with N/A in Inverted Ordering).}
  \label{table_BF_and_LL}
 \setlength{\tabcolsep}{6pt}
 \renewcommand{\arraystretch}{1.3}
\rotatebox[]{90}{
        \begin{tabular}{clrclcl}
  \hline
  \hline
                       &                                  &                                                 & \multicolumn{2}{l}{ \hspace{1mm} Normal MO  } & \multicolumn{2}{l}{ \hspace{1mm} Inverted MO} \\
        Reactor Constraint & Fit Constraint                   & Parameter                                       & HPD                                      & \hspace{2mm} 1$\sigma$ range                  & HPD    & \hspace{2mm} 1$\sigma$ range  \\
  \hline
  \ \\
        \multirow{8}{*}{None} 
                       & \multirow{4}{*}{Conditional}     & \(\delta_{\rm CP}(\pi)\)                        & 0.89                                          & [0.00,  0.24] $\cup$ [0.50, 1.15]           & 1.57         & [1.32, 1.78]               \\
                       &                                  & \(\sin^2\theta_{23}\)                           & 0.547                                         & [0.465, 0.509] $\cup$ [0.517, 0.562]        & 0.474        & [0.453, 0.537]             \\
                       &                                  & \(\Delta m^2_{32}(\times 10^{-3} \text{eV}^2)\) & 2.431                                         & [2.392, 2.467]                            & -2.476       & [-2.515, -2.440]           \\
                       &                                  & \(\sin^22\theta_{13}(\times 10^{-2})\)          & 8.5                                           & [7.1, 10.0]                               & 9.6          & [8.3, 11.2]                \\
                       \ \\
                       & \multirow{4}{*}{Non-Conditional} & \(\delta_{\rm CP}(\pi)\)                        & 0.89                                          & [0.00,  0.24] $\cup$ [0.50, 1.15]           & N/A          & [1.33, 1.78]               \\
                       &                                  & \(\sin^2\theta_{23}\)                           & 0.547                                         & [0.450, 0.576]                            & N/A          & [0.466, 0.482]             \\
                       &                                  & \(\Delta m^2_{32}(\times 10^{-3} \text{eV}^2)\) & 2.431                                         & [2.392, 2.480]                            & N/A          & [-2.491, -2.456]           \\
                       &                                  & \(\sin^22\theta_{13}(\times 10^{-2})\)          & 8.5                                           & [6.6, 10.6]                               & N/A          & [9.0, 10.3]                \\
        \ \\
        \hline
        \ \\
        \multirow{6}{*}{1D} 
                       & \multirow{3}{*}{Conditional}     & \(\delta_{\rm CP}(\pi)\)                        & 0.93                                          & [0.04, 0.29] $\cup$ [0.62, 1.14]            & 1.50         & [1.30, 1.69]               \\
                       &                                  & \(\sin^2\theta_{23}\)                           & 0.548                                         & [0.484, 0.566]                            & 0.545        & [0.489, 0.565]             \\
                       &                                  & \(\Delta m^2_{32}(\times 10^{-3} \text{eV}^2)\) & 2.431                                         & [2.397, 2.467]                            & -2.479       & [-2.515, -2.443]           \\
                       \ \\
                       & \multirow{3}{*}{Non-Conditional} & \(\delta_{\rm CP}(\pi)\)                        & 0.93                                          & [0.02, 0.34] $\cup$ [0.58, 1.15]            & N/A          & [1.37, 1.63]               \\
                       &                                  & \(\sin^2\theta_{23}\)                           & 0.548                                         & [0.459, 0.575]                            & N/A          & \(\emptyset\)              \\
                       &                                  & \(\Delta m^2_{32}(\times 10^{-3} \text{eV}^2)\) & 2.431                                         & [2.376, 2.485]                            & N/A          & [-2.480, -2.472]           \\
        \ \\
        \hline
        \ \\
        \multirow{6}{*}{2D} 
                       & \multirow{3}{*}{Conditional}     & \(\delta_{\rm CP}(\pi)\)                        & 0.92                                          & [0.09, 0.28] $\cup$ [0.59, 1.14]            & 1.55         & [1.33, 1.72]               \\
                       &                                  & \(\sin^2\theta_{23}\)                           & 0.550                                         & [0.473, 0.498] $\cup$ [0.508, 0.569]        & 0.554        & [0.472, 0.492] $\cup$ [0.510, 0.574] \\
                       &                                  & \(\Delta m^2_{32}(\times 10^{-3} \text{eV}^2)\) & 2.439                                         & [2.408, 2.472]                            & -2.497       & [-2.531, -2.467]           \\
                       \ \\
                       & \multirow{3}{*}{Non-Conditional} & \(\delta_{\rm CP}(\pi)\)                        & 0.92                                          & [0.05, 0.36] $\cup$ [0.53, 1.16]            & N/A          & \(\emptyset\)              \\
                       &                                  & \(\sin^2\theta_{23}\)                           & 0.550                                         & [0.470, 0.571]                            & N/A          & \(\emptyset\)              \\
                       &                                  & \(\Delta m^2_{32}(\times 10^{-3} \text{eV}^2)\) & 2.439                                         & [2.400, 2.477]                            & N/A          & \(\emptyset\)              \\
        \hline
  \hline
  \end{tabular}
}
  \end{table*}

\subsubsection{Goodness of fit} \label{sec:goodness_of_fit}

The posterior-predictive $p$-values~\cite{Gelman:1996} are obtained from the Bayesian MCMC fit to evaluate the goodness of the fit. 
The posterior-predictive $p$-values, split by FD data sample and reactor constraints used for $\theta_{13}$, are presented in Table~\ref{table:pvals}, with a good fit seen across all the data samples. The differences in goodness of fit between the fits with and without the external reactor constraints are minimal. The electron-antineutrino sample has the largest deviation from 0.5, and the new Low-energy electron-neutrino sample has the largest posterior-predictive $p$-value. These are the two FD data samples containing the fewest events.

\begin{table}[htpb]
    \centering
    \caption{Posterior-predictive $p$-values extracted from the Bayesian MCMC fit of the real data with different types of reactor constraint applied to $\theta_{13}$.}
    \label{table:pvals}
    \begin{tabular}{lcccccc}
      \hline
      \hline\\[-2.3ex]
      Constraint type & $\nu_\mu$ & $\nu_e$ & Low-energy& $\bar{\nu}_\mu$ & $\bar{\nu}_e$ & Total \\
      \hline\\[-2.3ex]
      Unconstrained & 0.59 & 0.57 & 0.72 & 0.40 & 0.17 & 0.48 \\
      1D Daya Bay    & 0.59 & 0.58 & 0.72 & 0.39 & 0.19 & 0.48 \\
      2D Daya Bay   & 0.58 & 0.58 & 0.72 & 0.40 & 0.18 & 0.49 \\
      \hline
      \hline
    \end{tabular}
  \end{table}

\subsubsection{Posterior Probability Densities}
\label{sec:posteriors}

\begin{figure*}[htb]
  \includegraphics[width=0.9\linewidth]{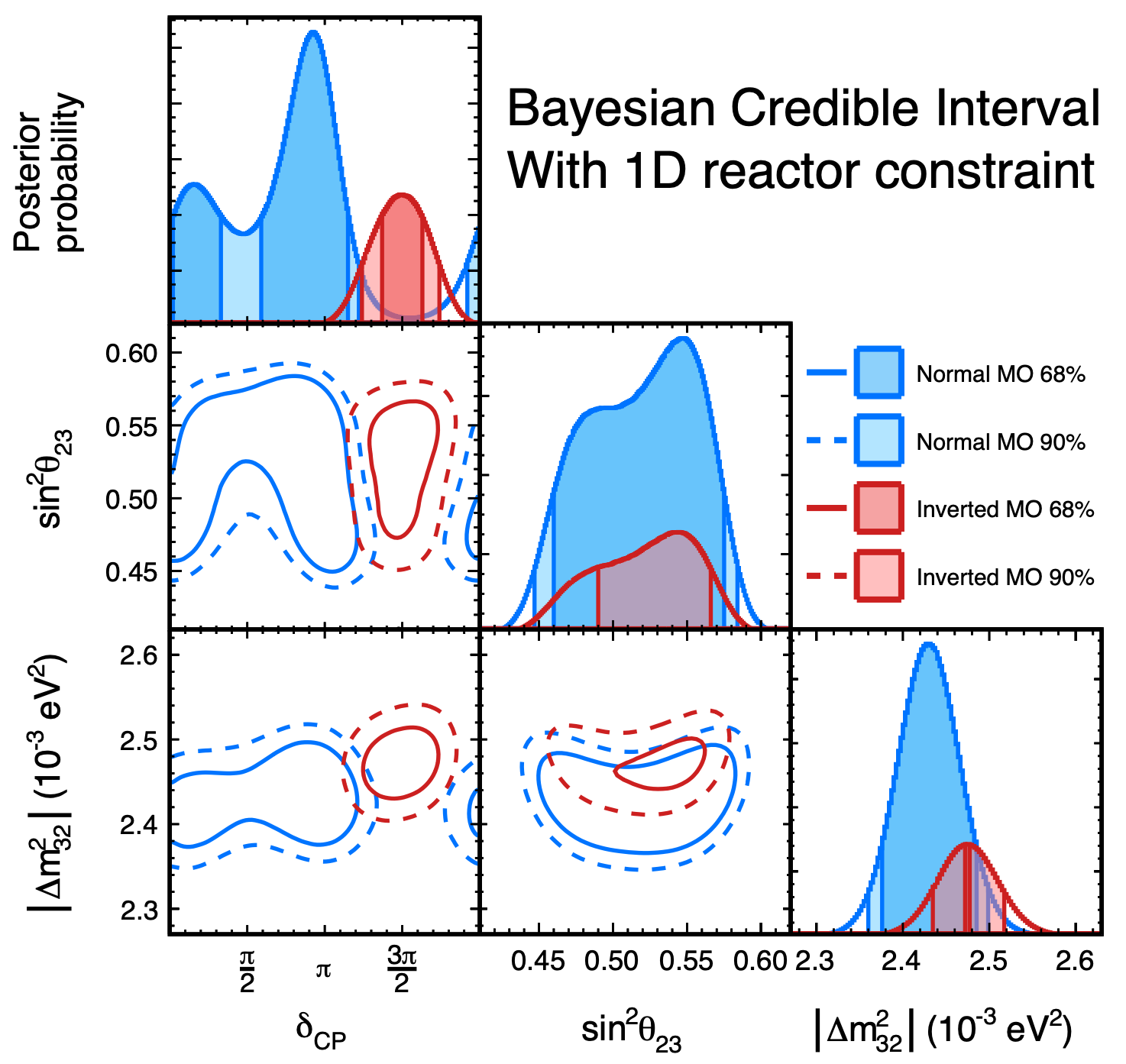}
  \caption{Marginalized posterior probability densities from a Bayesian fit with the 1D Daya Bay constraint on
    $\sin^2 2\theta_{13}$ applied. The two-dimensional posteriors have
    68\% and 90\% credible levels indicated. The one-dimensional posterior
    probability densities are normalized to 1, and have 68\% and 90\%
    credible intervals indicated in different shades. These plots are
    calculated by marginalizing over both MOs.}
\label{fig:triangle}
\end{figure*}

This section presents one- and two-dimensional marginalized posterior distributions for the measured oscillation parameters.
The 68\% and 90\% credible intervals for the marginalized posterior probability densities of $|\Delta m^2_{32}|$, $\sin^2\theta_{23}$, and $\delta_{\textrm{CP}}$, obtained using the 1D Daya Bay constraint on $\sin^2 2\theta_{13}$, are summarized in Fig.~\ref{fig:triangle}. All posteriors are marginalized jointly over both mass orderings.

\begin{figure*}[htb]
\includegraphics[width=0.6\linewidth]{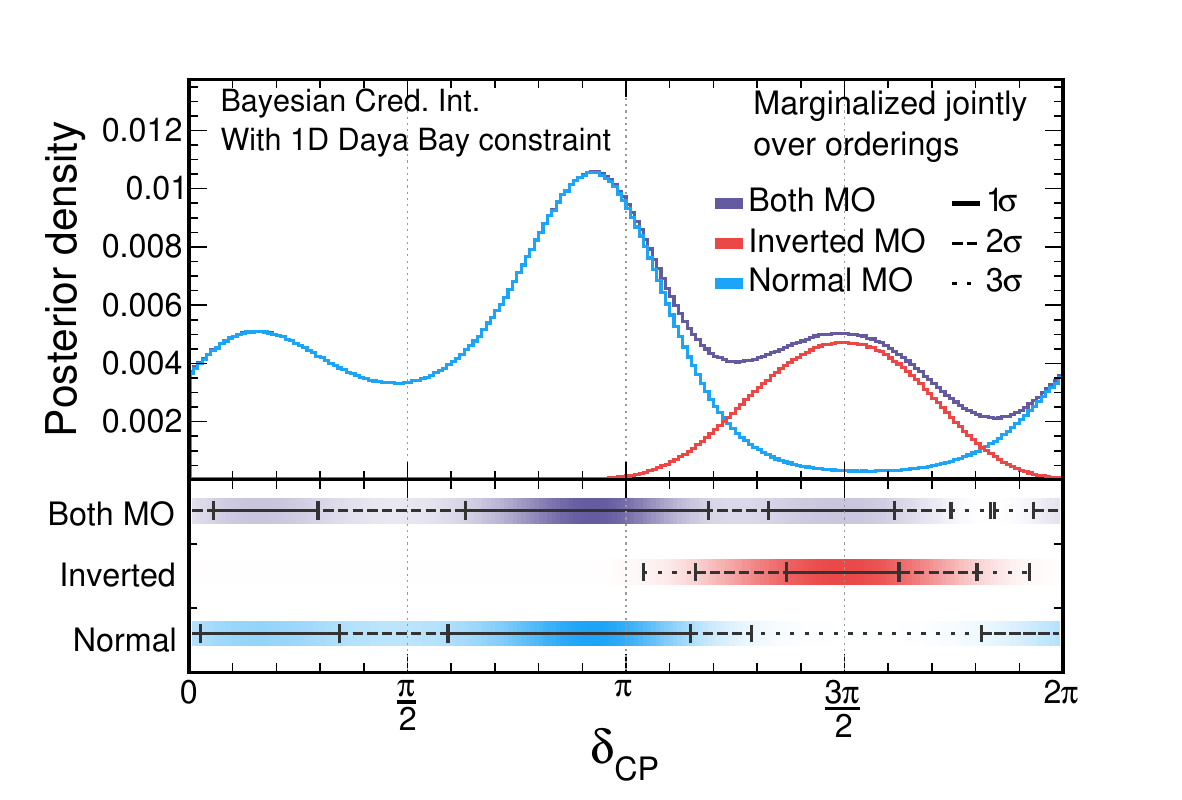}
  \caption{Posterior probability densities for $\delta_{\textrm{CP}}$
           marginalized jointly over the MOs, shown for both
           MO (purple), Inverted MO (red), and Normal MO (blue), with the 1, 2,
           and 3\,$\sigma$ intervals shown below. Posterior extracted from a
           fit with an external 1D constraint on $\theta_{13}$ from the Daya Bay
           experiment.}
\label{fig:1D_nonconditional_dcp with_bands_rcrw1D}
\end{figure*}

\begin{figure*}[htb]
\includegraphics[width=0.49\linewidth]{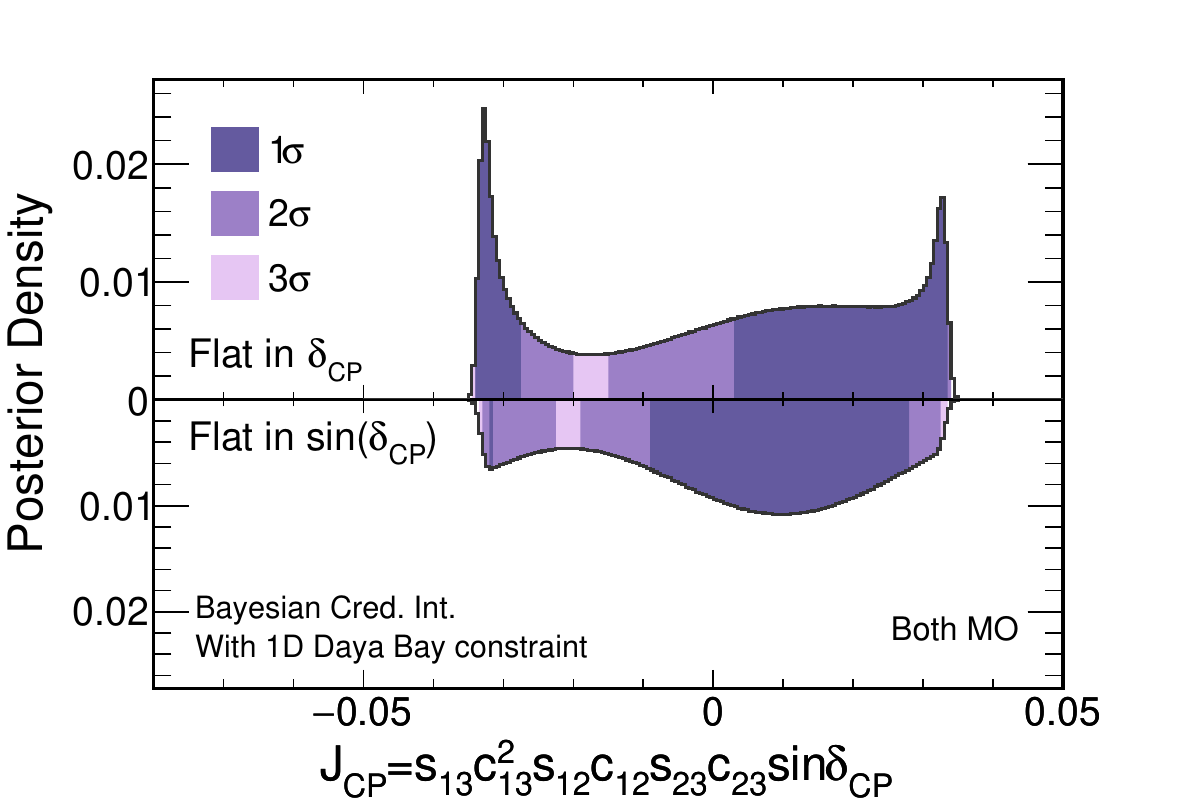}\\
\includegraphics[width=0.49\linewidth]{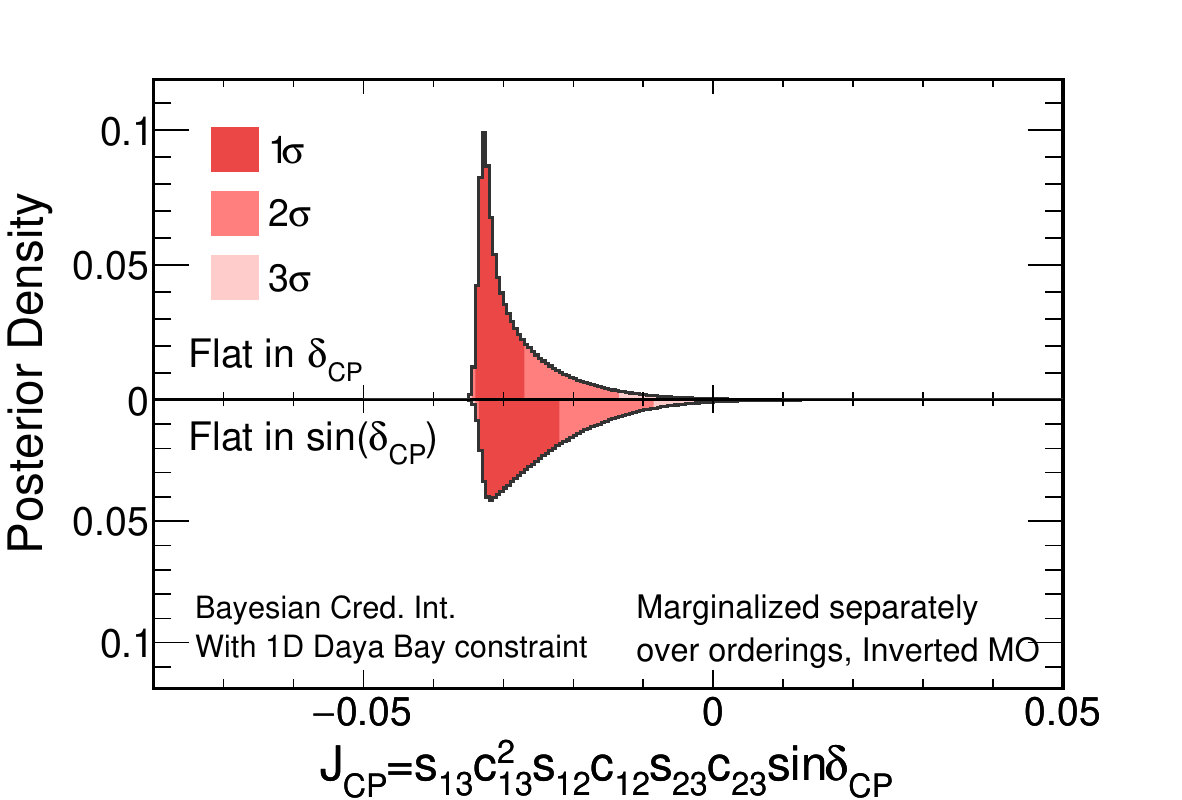}
\includegraphics[width=0.49\linewidth]{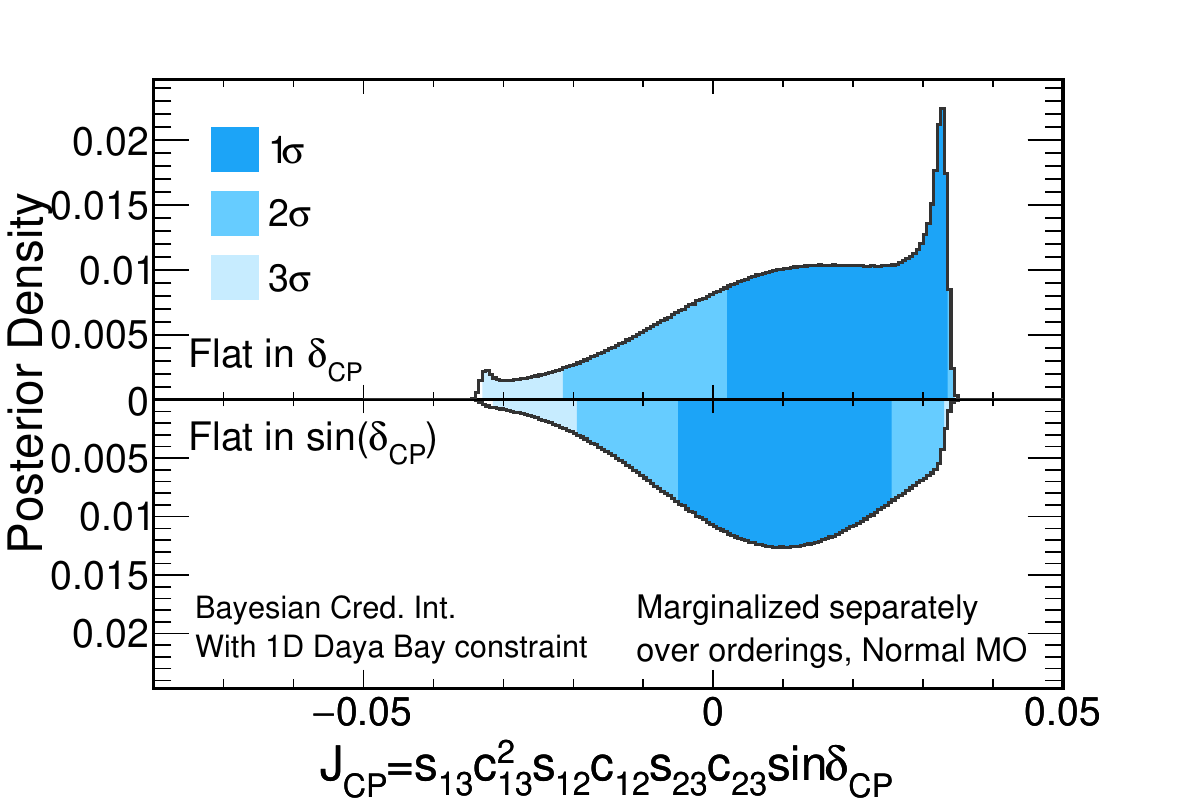}
  \caption{Posterior probability densities for the Jarlskog invariant,
           marginalized separately over each ordering, and shown for the both
           MO (purple, top), Inverted MO (red, left) and Normal MO (blue,
           right). On each plot, the top panel shows the posterior with a
           prior uniform in $\delta_{\textrm{CP}}$, and the prior uniform in
           $\sin(\delta_{\textrm{CP}})$ in the upside-down canvas.
           Jarlskog invariant of 0 represents CP conservation, and non-zero
           values represent CP violation. The posterior is extracted from a fit with
           an external 1D constraint on $\theta_{13}$ from the Daya Bay
           experiment.}
\label{fig:jarlskog}
\end{figure*}

\begin{figure*}[htb]
\includegraphics[width=0.49\linewidth]{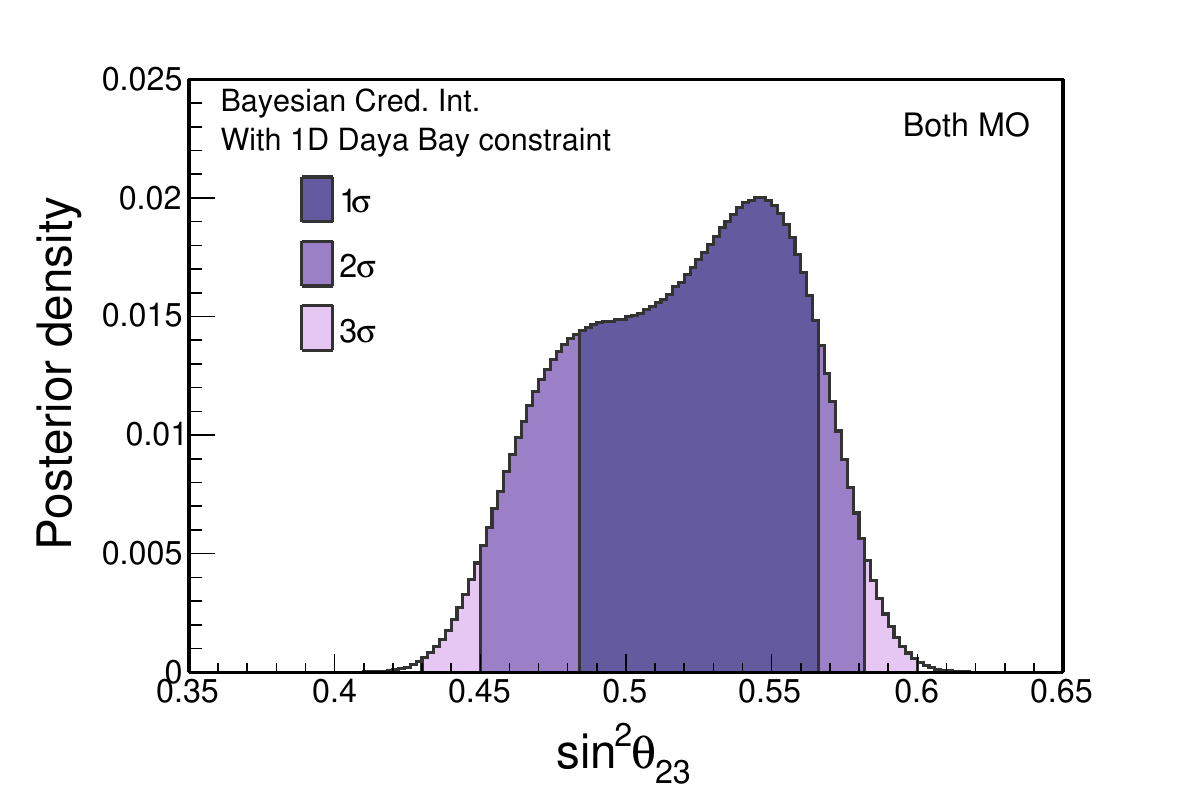}\\
\includegraphics[width=0.49\linewidth]{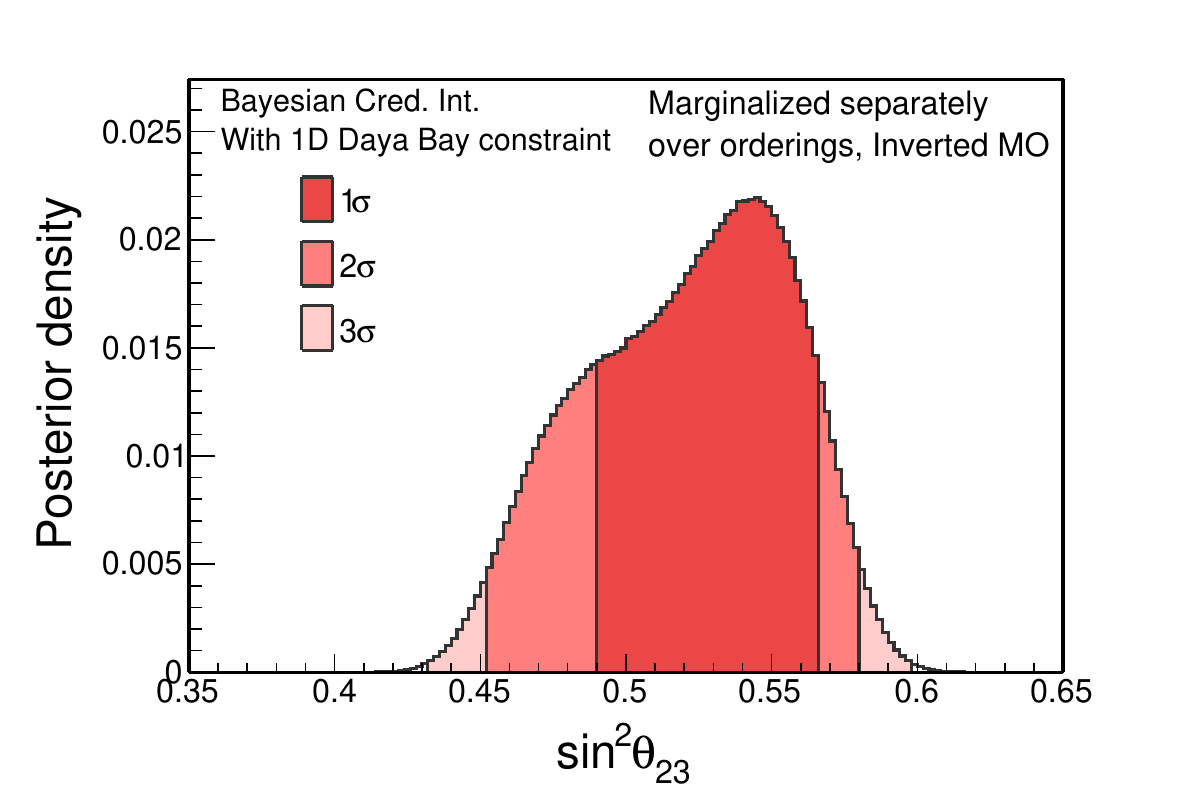}
\includegraphics[width=0.49\linewidth]{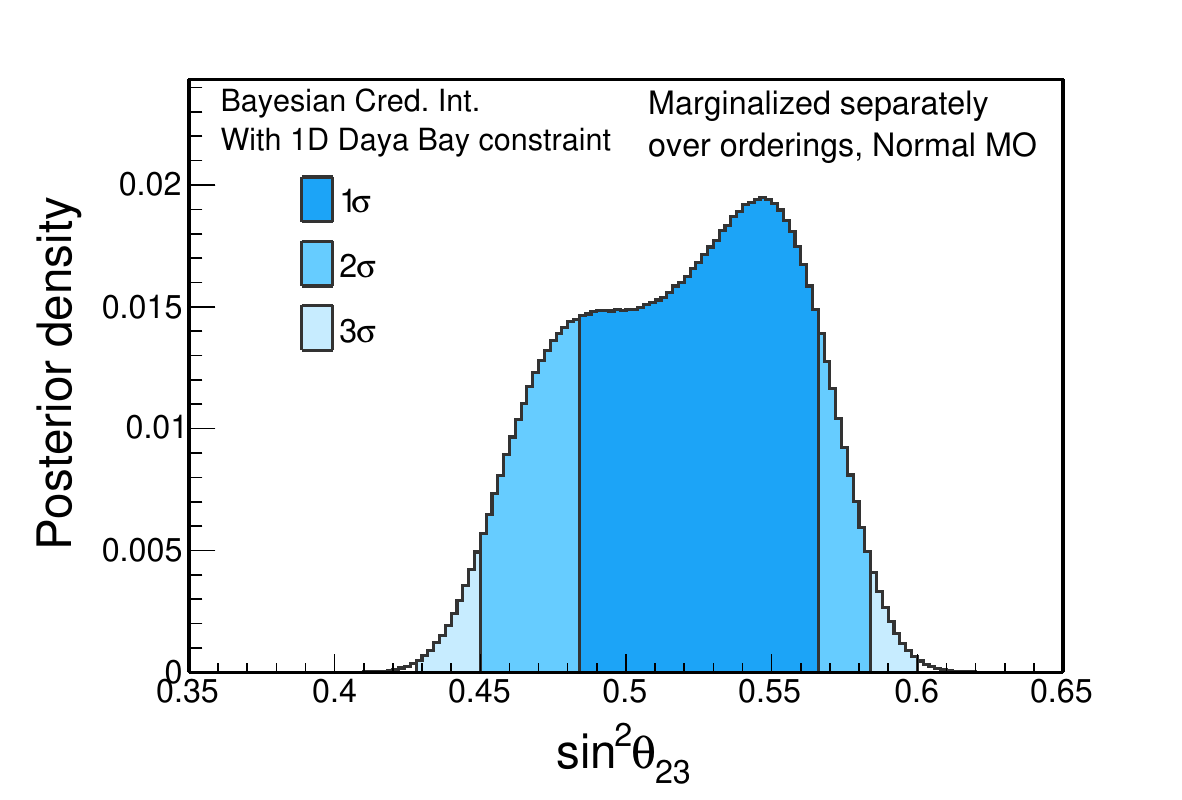}
  \caption{Posterior probability densities for $\sin^2\theta_{23}$
           marginalized separately over the MOs, shown for both
           MO (purple, top), Inverted MO (red, left), and Normal MO (blue,
           right). The posterior is from a fit with an external 1D
           constraint on $\theta_{13}$ from the Daya Bay experiment.}
\label{fig:1D_ssth23_rcrw1D}
\end{figure*}

One-dimensional constraints on the CP violation phase $\delta_{\textrm{CP}}$, marginalized jointly over both MOs, are presented in Fig.~\ref{fig:1D_nonconditional_dcp with_bands_rcrw1D}. The two MOs exhibit highest posterior probability densities in different regions of parameter space.
Credible intervals for the Jarlskog invariant, with priors uniform in $\delta_{\textrm{CP}}$ and in $\sin \delta_{\textrm{CP}}$, are shown in Fig.~\ref{fig:jarlskog}. Probabilities are marginalized over both MOs, with individual MO plots also provided. In the Inverted ordering, which is less favored by the NOvA analysis, the majority of the probability density is away from the CP-conserving point at $J_{\textrm{CP}}=0$.
The posterior probability density for $\sin^2\theta_{23}$ is shown in Fig.~\ref{fig:1D_ssth23_rcrw1D}, with probabilities shown for both MO and marginalized separately for each MO. The favored parameter regions are compatible in all cases.

\begin{figure*}[htb]
\includegraphics[width=0.49\linewidth]{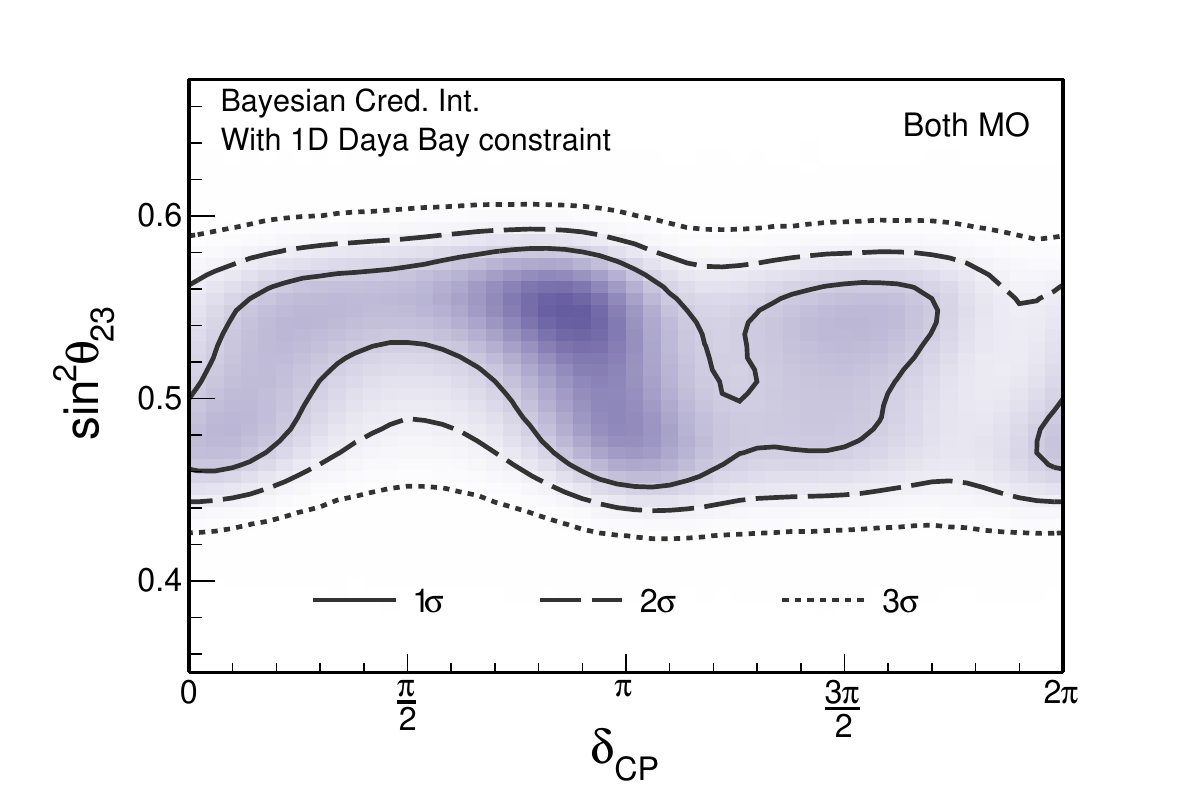} \\
\includegraphics[width=0.49\linewidth]{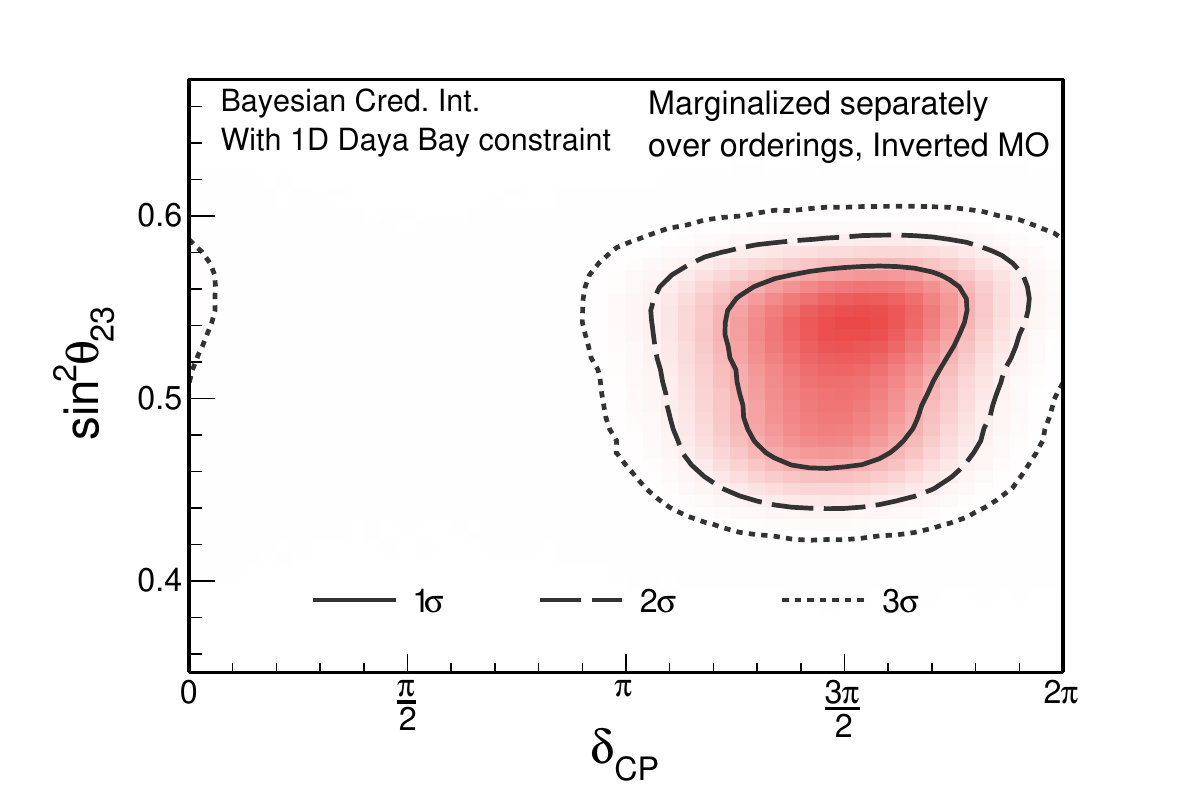}
\includegraphics[width=0.49\linewidth]{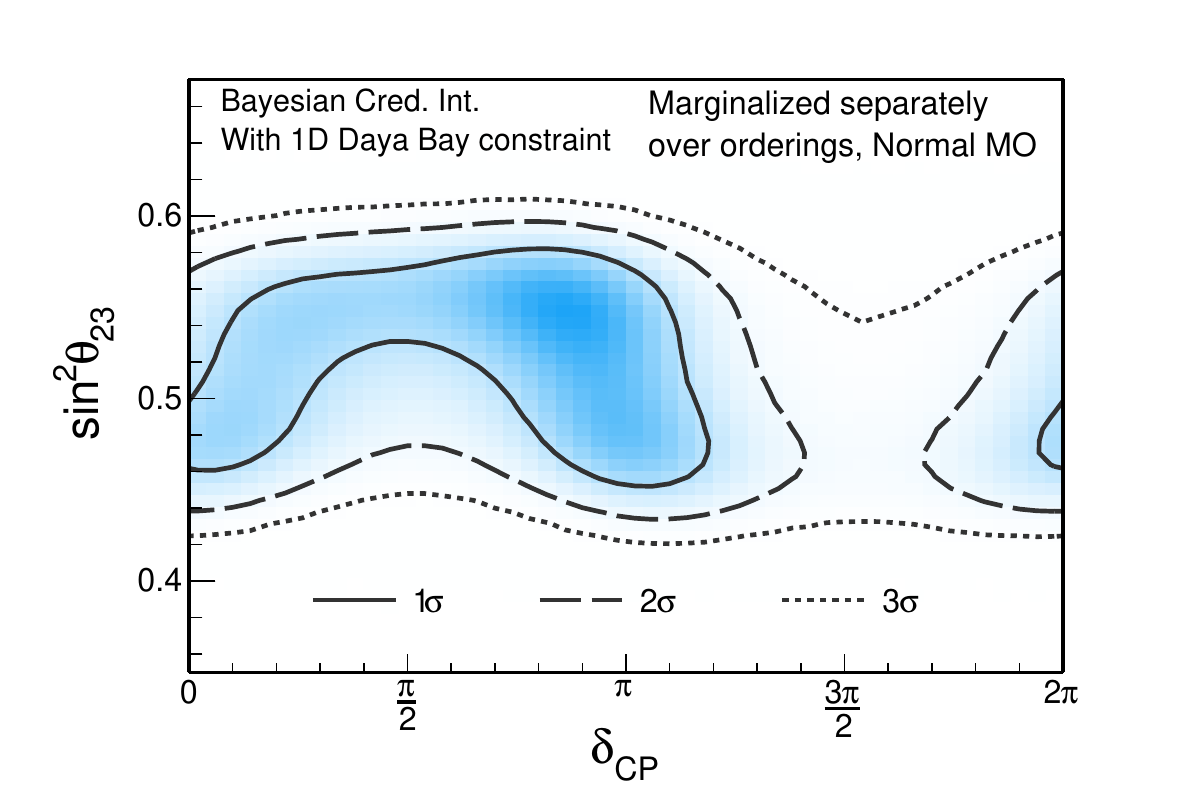}
  \caption{Posterior probability density in the plane $\sin^2\theta_{23}$
           vs $\delta_{\textrm{CP}}$, marginalized separately over
           the MOs, shown for both MO (purple, top),
           Inverted MO (red, left) and Normal MO (blue, right). Posteriors
           are from the fits to NOvA data with an external constraint
           on $\theta_{13}$ from the Daya Bay experiment.}
\label{fig:2D_dcp_ssth23_rcrw1D}
\end{figure*}

\begin{figure*}[htb]
\includegraphics[width=0.49\linewidth]{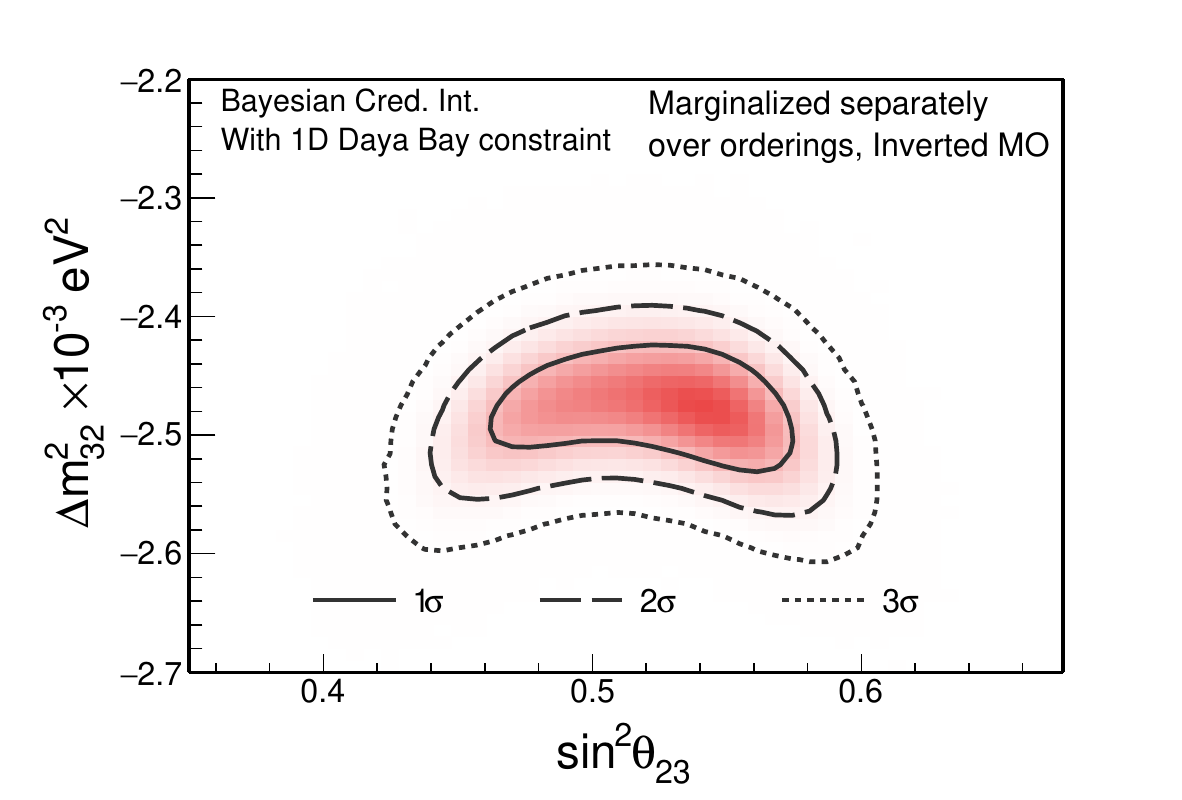}
\includegraphics[width=0.49\linewidth]{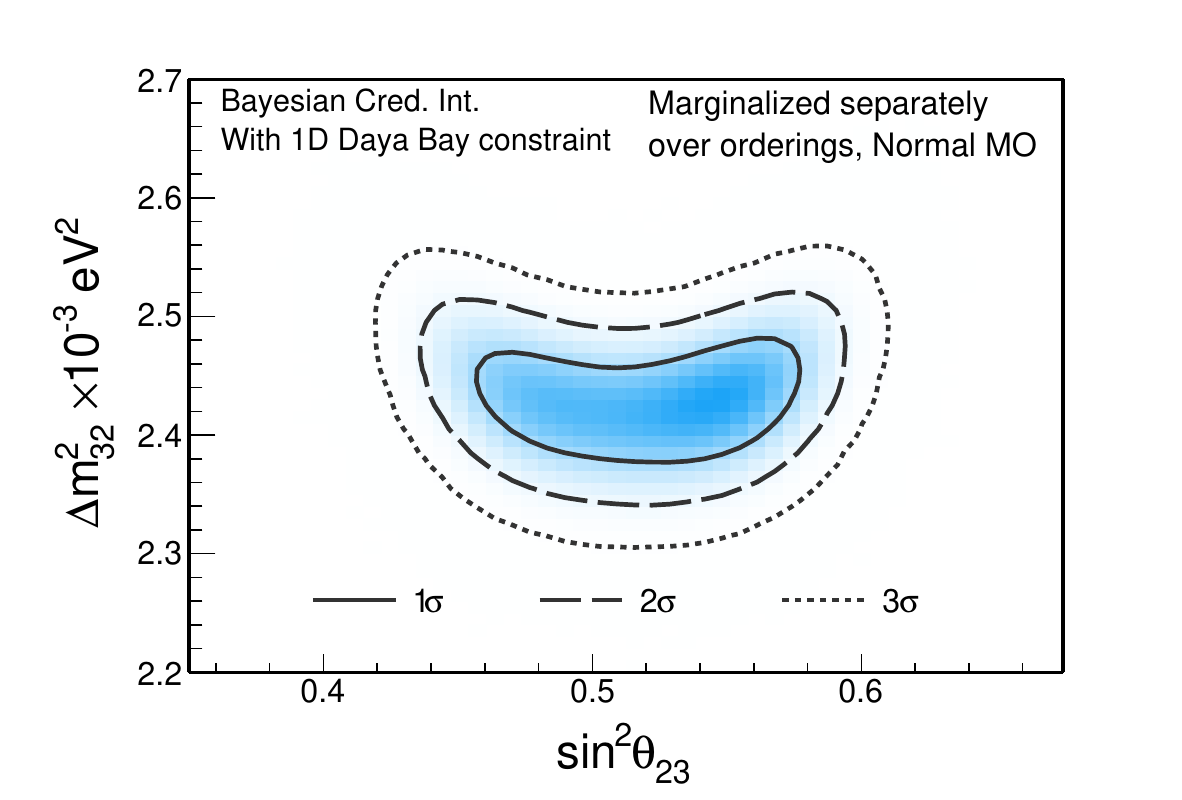}
  \caption{Posterior probability density in the plane $\Delta m^2_{32}$
           vs $\sin^2\theta_{23}$, marginalized separately over the
           MOs, shown for the Inverted (red, left) and Normal
           MO (blue, right). The posteriors are from the fits to NOvA
           data with an external constraint on $\theta_{13}$ from the Daya Bay
           experiment.}
\label{fig:2D_ssth23_dm32_rcrw1D}
\end{figure*}

Figure~\ref{fig:2D_dcp_ssth23_rcrw1D} presents the two-dimensional posterior distribution in the plane of $\sin^2\theta_{23}$ versus $\delta_{\textrm{CP}}$, incorporating the one-dimensional Daya Bay constraint on $\theta_{13}$. Three versions are shown: marginalized jointly over both MOs, or separately for each of the two MO. The NOvA data prefer different $\delta_{\textrm{CP}}$ regions for the two MOs.
Posterior probability densities in the plane of $\Delta m^2_{32}$ versus $\sin^2\theta_{23}$ are shown in Fig.~\ref{fig:2D_ssth23_dm32_rcrw1D}, with the MOs marginalized separately. The favored region is consistent with maximal mixing, with no strong octant preference.

\begin{figure*}[htb]
\includegraphics[width=0.49\linewidth]{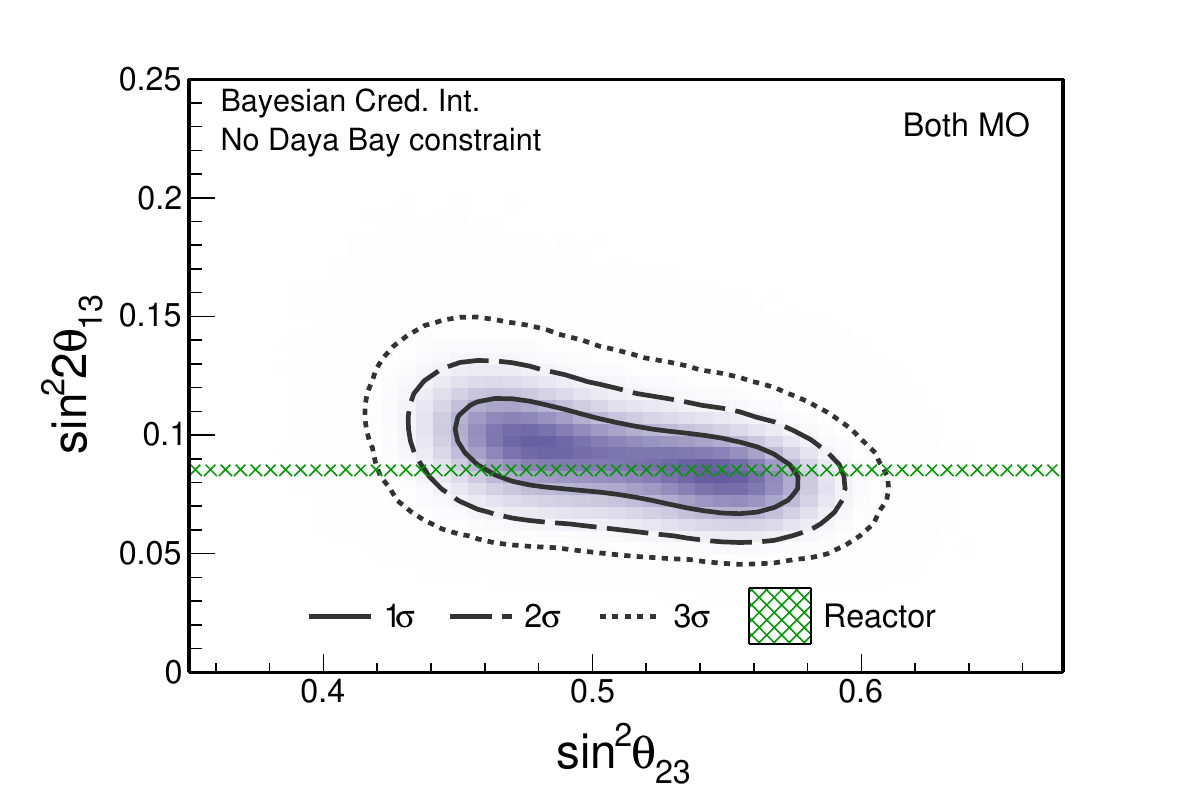}\\
\includegraphics[width=0.49\linewidth]{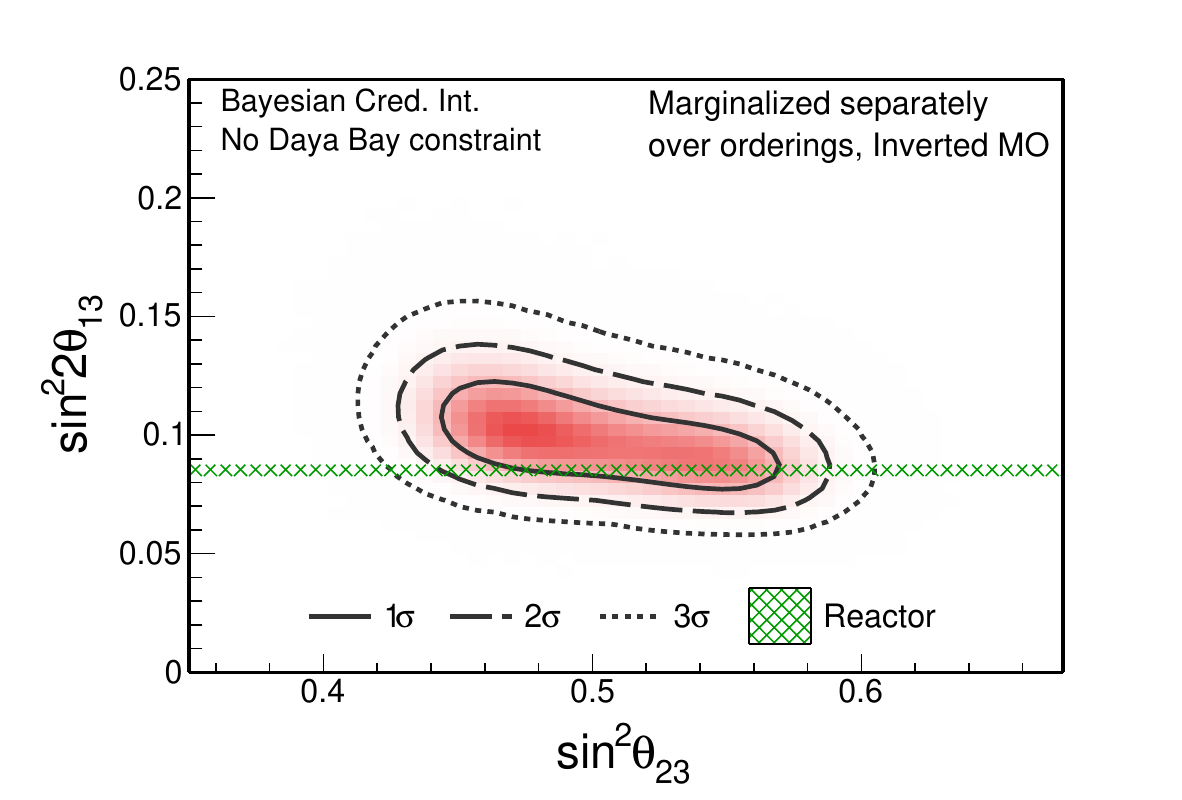}
\includegraphics[width=0.49\linewidth]{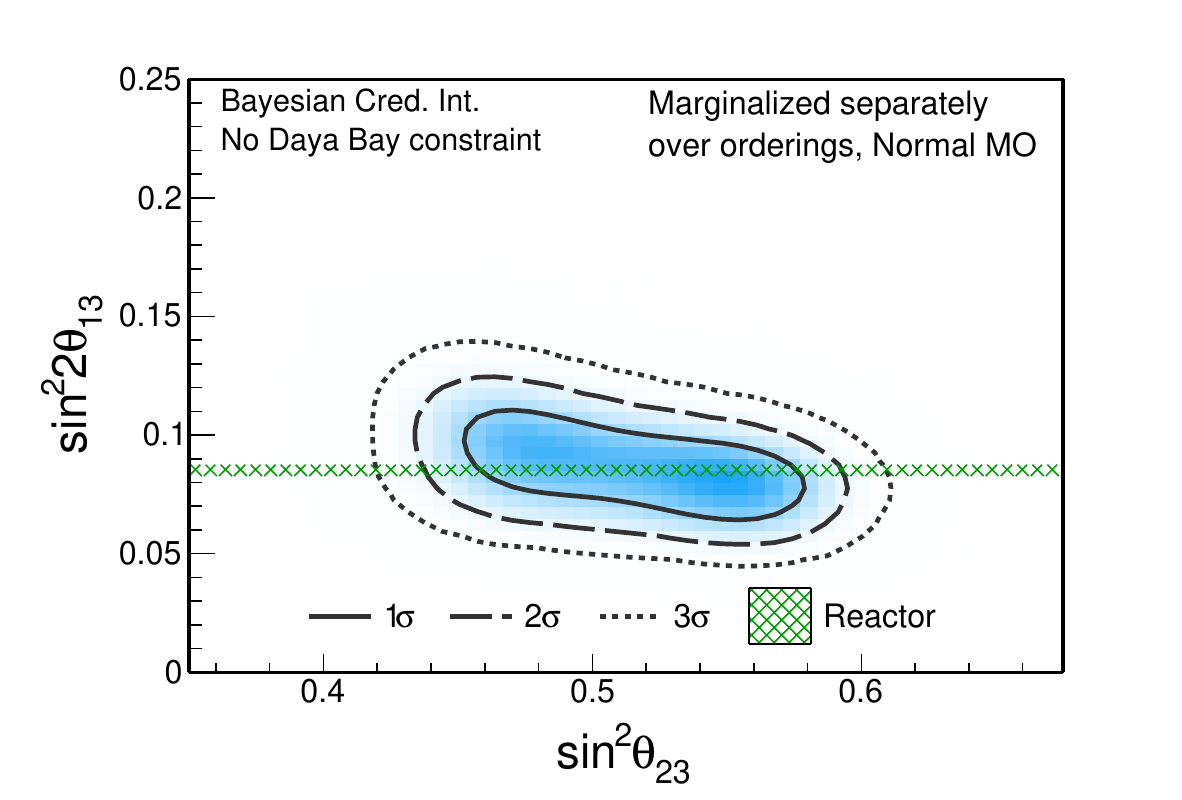}
  \caption{Posterior probability density in the plane $\sin^22\theta_{13}$
           vs $\sin^2\theta_{23}$, marginalized jointly over
           the MOs, shown for both MO (purple, top),
           Inverted MO (red, left) and Normal MO (blue, right). Posteriors
           are from the fits to NOvA data without any external
           constraints, with results from the Daya Bay experiment overlaid as
           green 1\,$\sigma$ band~\cite{DayaBay:2022orm}.}
\label{fig:2D_ssth23_ss2th13_BO}
\end{figure*}

Figure~\ref{fig:2D_ssth23_ss2th13_BO} shows the strong correlations between $\sin^22\theta_{13}$ and $\sin^2\theta_{23}$ in long-baseline oscillation measurements, along with a comparison to the reactor neutrino measurement of $\theta_{13}$, which places a strong constraint on this parameter. The consistency between the measurements supports its use as a constraint in this analysis and highlights the effectiveness of the three-flavor description of neutrinos.

\FloatBarrier
\subsubsection{Comparisons} \label{sec:comparisons}

\begin{figure}[!htb]
\includegraphics[width=0.49\linewidth]{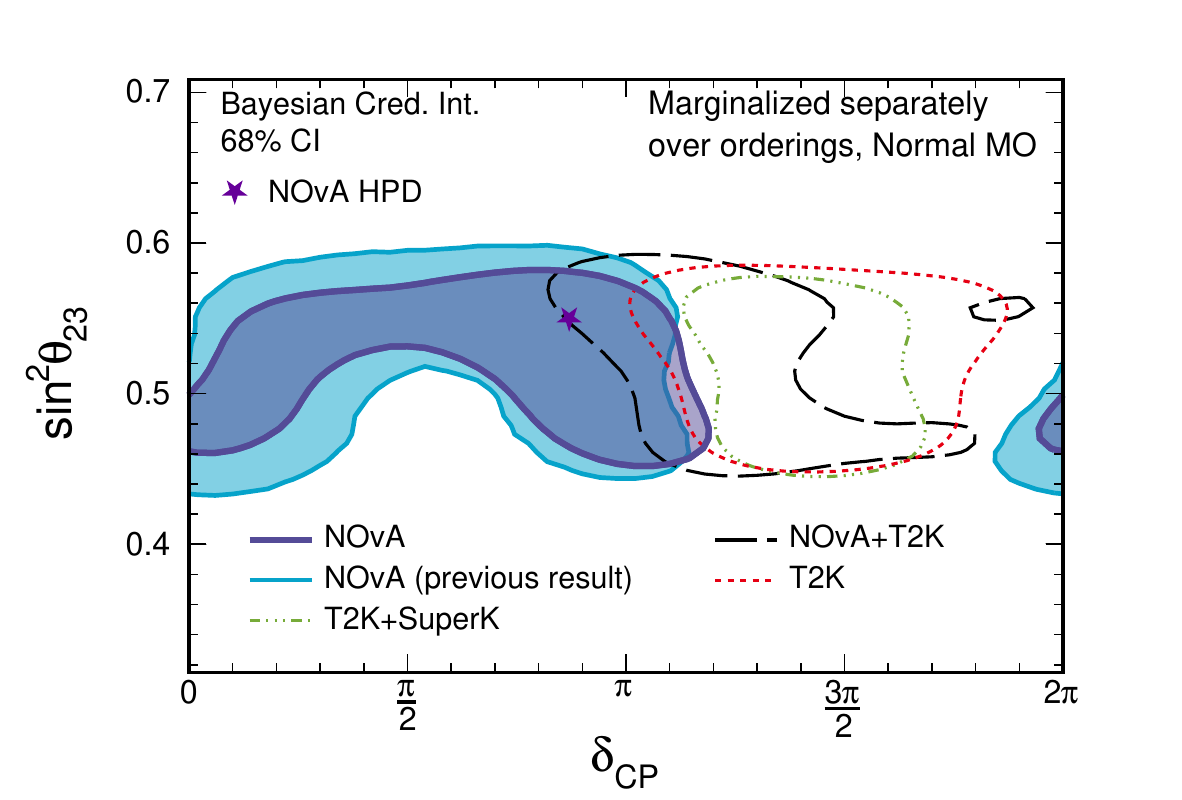}
\includegraphics[width=0.49\linewidth]{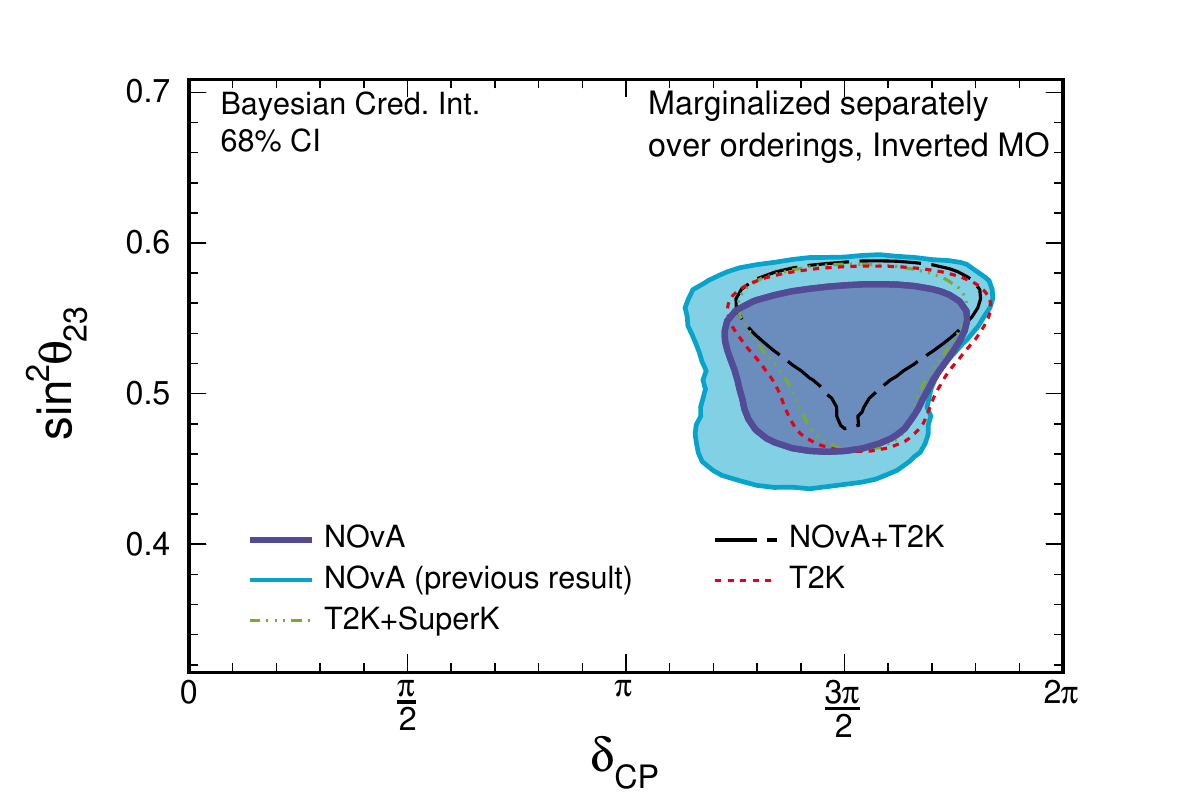}
  \caption{New NOvA results for $\delta_{\textrm{CP}}$ -- $\sin^2\theta_{23}$ in the
           Normal (left) and Inverted (right) MOs, compared with previous NOvA results~\cite{NOvA:2021nfi, NOvA:2023iam} and results from other
           experiments~\cite{T2K:2024wfn, Super-Kamiokande:2023ahc, t2k_collaboration_2022_6908532}, including the 2024 joint NOvA-T2K analysis~\cite{novat2k_nature2025}. All
           90\% credible intervals are from Bayesian analyses. }
\label{fig:plot_2D_dcp_ssth23_rcrw1D_nh_comparison_w_other_experiments_with_friends}
\end{figure}

This section provides a global comparison of NOvA measurements with those from other neutrino experiments.
Figure~\ref{fig:plot_2D_dcp_ssth23_rcrw1D_nh_comparison_w_other_experiments_with_friends} compares the measurements of $\sin^2\theta_{23}$ versus $\delta_{\textrm{CP}}$ from NOvA with corresponding results from other experiments for the Normal (left) and Inverted (right) MOs.
T2K data tend to favor values of $\delta_{\textrm{CP}}$ near $3\pi/2$ in the Normal MO. However, the 68\% credible intervals of NOvA and T2K show substantial overlap, indicating that current data are insufficient to draw strong conclusions about the value of $\delta_{\textrm{CP}}$, even when combining results across experiments. 

\begin{figure}[!htb]
\includegraphics[width=0.49\linewidth]{figures/main/plot_2D_ssth23_dm32_rcrw1D_nh_comparison_w_other_experiments_with_friends_v8.pdf}
\includegraphics[width=0.49\linewidth]{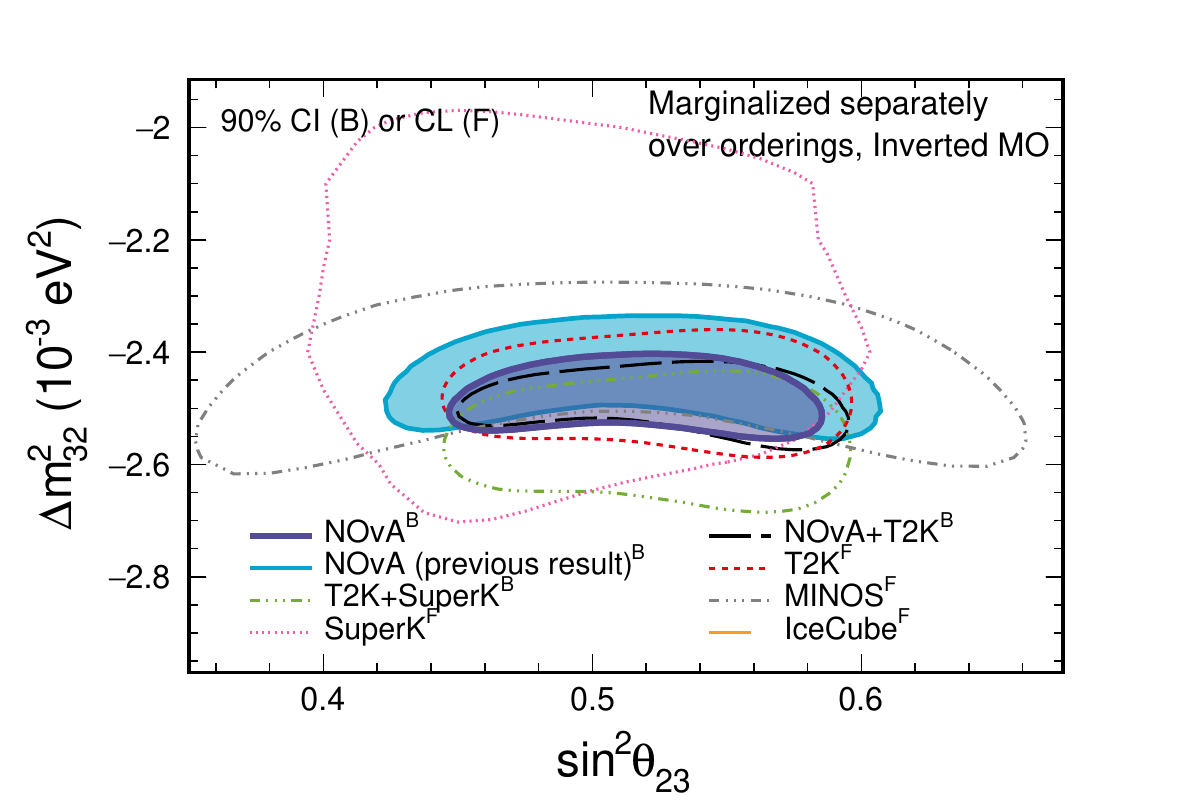}
  \caption{New NOvA results for $\Delta m^2_{32}$ -- $\sin^2\theta_{23}$ in the
           Normal (left) and Inverted (right) MOs, compared with previous NOvA results~\cite{NOvA:2021nfi, NOvA:2023iam} and results from other experiments~\cite{T2K:2024wfn,  IceCube:2024xjj, t2k_collaboration_2022_6908532, MINOS:2020llm, Super-Kamiokande:2023ahc}, including
           the 2024 joint NOvA-T2K analysis~\cite{novat2k_nature2025}. Contours labeled B are from Bayesian analyses, while those labeled F are from frequentist analyses, used when Bayesian results were not available.}
\label{fig:plot_2D_ssth23_dm32_rcrw1D_nh_comparison_w_other_experiments_with_friends}
\end{figure}

A comparison of the NOvA $\Delta m^2_{32}$ vs $\sin^2\theta_{23}$ measurement
with similar measurements from other experiments in the Normal MO is shown in Fig.~\ref{fig:with_friends}. For completeness,
Fig.~\ref{fig:plot_2D_ssth23_dm32_rcrw1D_nh_comparison_w_other_experiments_with_friends}
shows the same plot together with a version in the Inverted MO.

\begin{figure}[htb]
  \includegraphics[width=\largefigwidth]{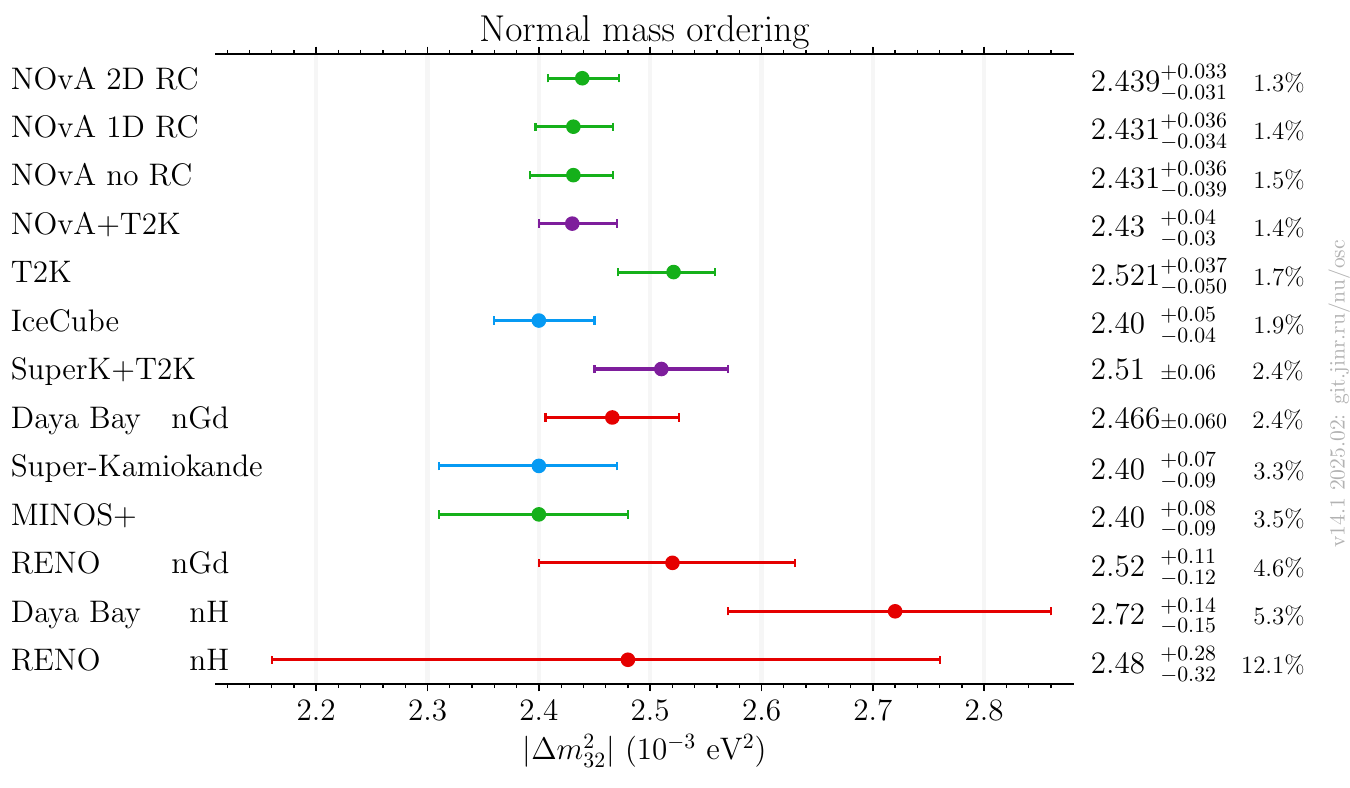}
  \includegraphics[width=\largefigwidth]{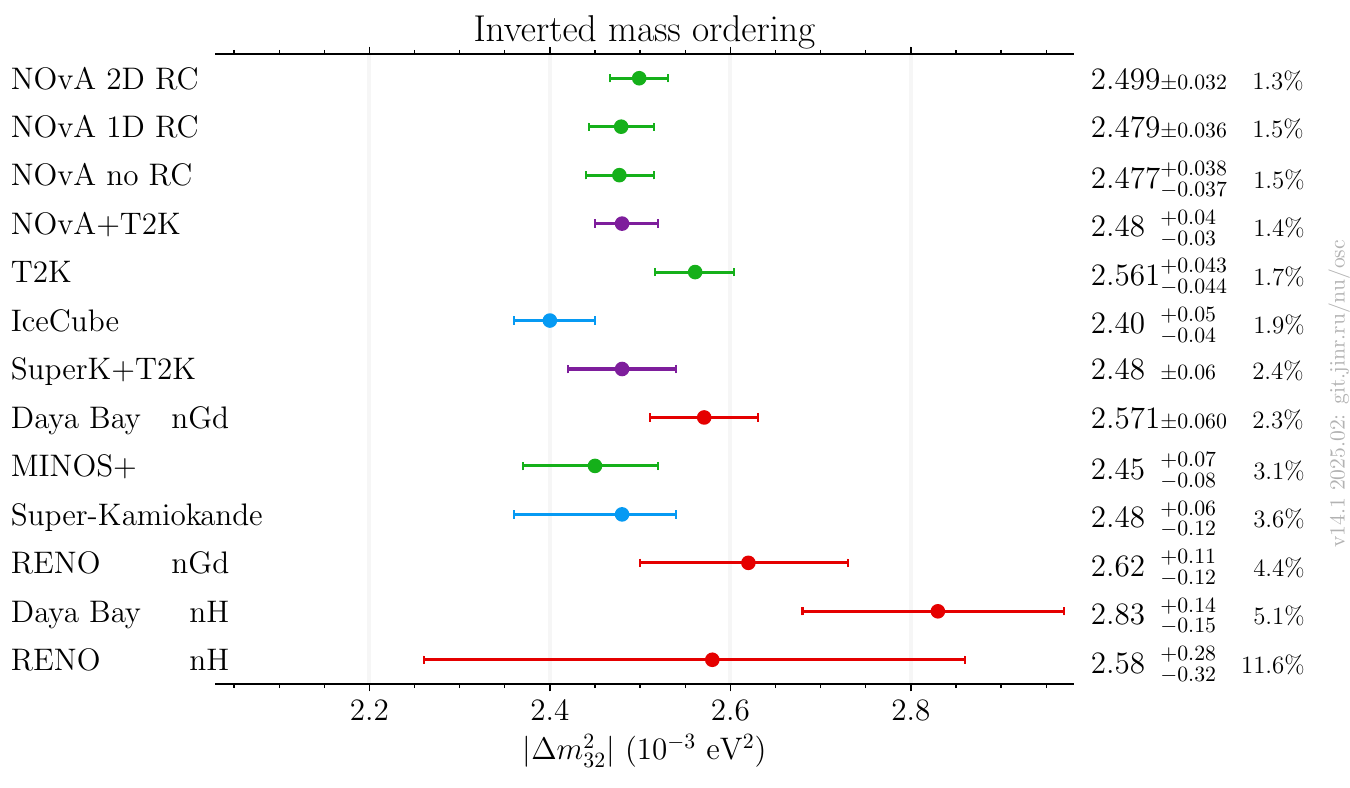}
  \caption{Comparison of central values and 1\,$\sigma$ intervals for
           absolute value of $\Delta m^2_{32}$ in the Normal
           MO (top) and Inverted MO (bottom) from various accelerator (green), reactor (red),
           atmospheric (blue) experiments, and the two joint fits (purple). Sources for the results from top to bottom starting with the fourth line are as follows: \cite{t2k2024, IceCube:2024xjj, T2K:2024wfn, DayaBay:2022orm, Super-Kamiokande:2023ahc, MINOS:2020llm, RENO:2024msr, DayaBay:2024hrv, reno_nH}.
           Comparisons for other oscillation
           parameters, can be found in~\cite{jinr:osc}.}
\label{fig:dm32_v11_latest}
\end{figure}

Figure~\ref{fig:dm32_v11_latest} shows the central values and 1$\sigma$ credible intervals of $\Delta m^2_{32}$ from NOvA alongside recent results from accelerator (green), atmospheric (blue), and reactor (red) experiments, as well as combined fits (purple).
Agreement among experiments is stronger in the Normal MO, contributing to the increased preference for this ordering in the NOvA analysis when the 2D reactor constraint from Daya Bay is included.

\clearpage
\FloatBarrier
\subsubsection{Mass ordering and $\theta_{23}$ octant preferences}
\label{sec:mass_ordering_octant}

Table~\ref{tab:bayesfactors} shows NOvA's preferences for the Normal MO and the upper octant of $\theta_{23}$, quantified using  Bayes factors and
the integrated posterior probabilities associated with these hypotheses.
Results are presented for three treatments of external reactor constraint: no constraint, and the 1D and 2D Daya Bay constraints, with no strong preferences are found for either the mass ordering or octant of $\theta_{23}$. Guidance on the traditional interpretation of Bayes factors can be found in~\cite{KassAndRaftery:1995,Jeffreys:1961}.

\begin{figure*}[!htb]
  \includegraphics[width=1\linewidth]{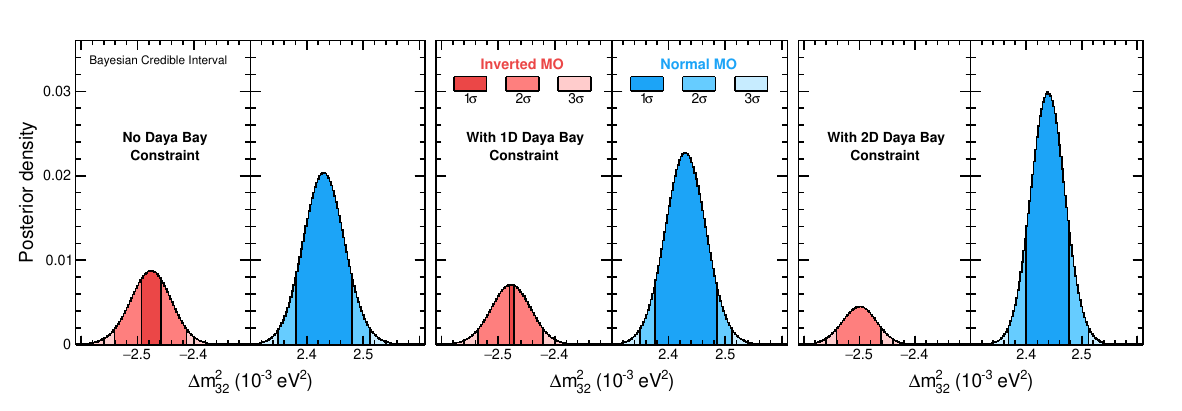}
\caption{Marginalized posterior probability densities for $\Delta m^2_{32}$ from a fit to data with different options of the Daya Bay constraint  applied (left panels  without any constraint, middle panels with $\theta_{13}$ constraint, right panels  with $\theta_{13}$--$\Delta m^2_{32}$ constraint).} 
\label{fig:MO}
\end{figure*}

Figure~\ref{fig:MO} further illustrates how different constraints on $\theta_{13}$ and $\Delta m^2_{32}$ from Daya Bay affect the MO preference, presenting the 1D posterior probabilities for $\Delta m^2_{32}$ in both Normal and Inverted orderings across the three choices on external constraint.

\begin{table}[!htb]
  \centering
  \caption{Bayes factors, with percentage posterior probability preference reported in parentheses, illustrating the preference for the Normal MO over the Inverted MO (top row) and for the upper octant over the lower octant (bottom row).Values are derived from fits using no external constraint on $\theta_{13}$, 1D and 2D constraints from Daya Bay.}
  \label{tab:bayesfactors}
  \begin{tabular}{l c c c}
    \hline
    \hline\\[-2.3ex]
     Bayes Factors & Unconstrained & 1D Daya Bay & 2D Daya Bay \\
    \hline\\[-2.3ex]
     Normal Ordering Preference & 2.4 (70\%) & 3.3 (77\%) & 6.6 (87\%) \\
     Upper Octant Preference & 1.3 (56\%) & 2.1 (68\%) & 2.0 (66\%) \\
    \hline
    \hline
  \end{tabular}
\end{table}

\FloatBarrier

\clearpage

\subsection{Frequentist results} \label{sec:freq_results}

NOvA has employed both frequentist and, more recently, Bayesian statistical frameworks to extract oscillation parameters.
The tools that facilitate NOvA's Bayesian treatment, as presented in the main letter, are relatively new and were first exercised in~\cite{NOvA:2023iam} to re-analyze the dataset from~\cite{NOvA:2021nfi}. All previous NOvA oscillation results, including those in the original analysis of that dataset,  used a frequentist approach to inference. In this framework, a minimization of the Poisson log-likelihood ratio between simulation and data is performed using Minuit~\cite{James:1994vla}. The Profiled Feldman-Cousins Method~\cite{Feldman:1997qc,NOvA:2022wnj} is then employed to correct confidence intervals to ensure proper coverage.

Here we present updated frequentist results based on the latest dataset, provided for complementarity with the Bayesian analysis in the main text.
A summary of the frequentist fits using the 1D Daya Bay external constraint on  $\sin^2\theta_{13}$ can be found in Table~\ref{table_freq_BF}. The Inverted MO rejection significance is 1.41\,$\sigma$, which corresponds to a p-value of 0.16.
The 2D Daya Bay constraint on $\sin^2\theta_{13}\textrm{--}\Delta m^2_{32}$ yields an Inverted MO rejection significance of 1.6\,$\sigma$ (p-value of 0.11).

\begin{table*}[!htbp]
\centering
  \caption{A summary of NOvA's constraints on oscillation parameters using a frequentist approach to inference. The results using two different fit constraints are shown. In the conditional fit, $\Delta m^2_{32}$ is restricted to be either positive or negative, corresponding to the Normal or Inverted MO, respectively. In the non-conditional fit, $\Delta m^2_{32}$ is allowed to take either sign. In this fit case the global Best Fit is denoted and N/A is used in Inverted MO to mark that this case is not preferred.}
  \label{table_freq_BF}
 
  \begin{tabular}{clrclcl}
  \hline
  \hline
                       &                                  &                                                 & \multicolumn{2}{l}{ \hspace{1mm} Normal MO  } & \multicolumn{2}{l}{ \hspace{1mm} Inverted MO} \\
        Reactor Constraint & Fit Constraint                   & Parameter                                       & Best Fit                                      & \hspace{2mm} 1$\sigma$ range                  & Best Fit     & \hspace{2mm} 1$\sigma$ range \\
  \hline
  \hline
        \multirow{2}{*}{1D} 
                       & \multirow{3}{*}{Conditional}     & \(\delta_{\rm CP}(\pi)\)                        & 0.87                                          & [-0.03, 0.16] $\cup$ [0.58, 1.17]               & 1.53         & [1.26, 1.76] \\
                       &                                  & \(\sin^2\theta_{23}\)                           & 0.547                                         & [0.482, 0.488] $\cup$ [0.497, 0.566]            & 0.541        & [0.477, 0.561] \\
                       &                                  & \(\Delta m^2_{32}(\times 10^{-3} \text{eV}^2)\) & 2.441                                         & [2.410, 2.476]                            & -2.481       & [-2.515, -2.447]  \\
                       & \multirow{3}{*}{Non-Conditional} & \(\delta_{\rm CP}(\pi)\)                        & 0.87                                          & [-0.04, 0.15] $\cup$ [0.58, 1.17]               & N/A           & \(\emptyset\)  \\
                       &                                  & \(\sin^2\theta_{23}\)                           & 0.547                                         & [0.483, 0.570]                            & N/A             & \(\emptyset\) \\
                       &                                  & \(\Delta m^2_{32}(\times 10^{-3} \text{eV}^2)\) & 2.441                                         & [2.407, 2.480]                            & N/A             & \(\emptyset\) \\
        \hline
        \hline
  \end{tabular}

\end{table*}

\clearpage

\end{document}